\documentclass[twocolumn]{aastex7}

\newcommand{\Msun}{M$_{\odot}$}
\newcommand{\Mbh}{$M_{\rm BH}$}

\newcommand{\Lsun}{L$_\odot$}

\newcommand{\ml}{\emph{M/L}}

\newcommand{\hst}{HST}

\newcommand{\kms}{km~s$^{-1}$}

\newcommand{\go}{G$_{\rm outer}$}

\newcommand{\cotwo}{$^{12} $CO(2$-$1)}
\newcommand{\vsys}{\ensuremath{v_\mathrm{sys}}}
\newcommand{\voff}{\ensuremath{v_\mathrm{off}}}

\newcommand{\coone}{$^{12} $CO(1$-$0)}

\newcommand{\kinms}{\texttt{KinMS}}
\newcommand{\skysampler}{\texttt{SkySampler}}
\newcommand{\dynesty}{\texttt{dynesty}}
\newcommand{\kinemetry}{\texttt{kinemetry}}
\newcommand{\galfit}{\texttt{GALFIT}}
\newcommand{\casa}{\texttt{CASA}}

\hypersetup{colorlinks, linkcolor=blue, citecolor=blue, urlcolor=blue}

\usepackage{multirow}
\usepackage{footnote}
\usepackage{newtxtext, newtxmath}
\usepackage{graphicx}	
\usepackage{amsmath}	
\usepackage{lineno}
\usepackage{caption}
\usepackage{xcolor}
\usepackage{rotating}
\usepackage{soul,xcolor}
\linenumbers

\graphicspath{{./}{figures/}}
\received{May 5, 2026}
\revised{\today}
\submitjournal{AAS Journals}

\shorttitle{Molecular-based SMBH measurement of NGC~315}
\shortauthors{D.\ D.\ Nguyen et al.}

\begin{document}
\title{ALMA $^{12}$CO(2–1) Gas Dynamics in NGC 315: A Multi-Method Benchmark for Supermassive Black Hole Mass Measurement}

\correspondingauthor{Dieu D.\ Nguyen} \email{dieun@umich.edu}
\author[0000-0002-5678-1008]{Dieu D.\ Nguyen}
\email{dieun@umich.edu}
\affiliation{Department of Astronomy, University of Michigan, 1085 South University Avenue, Ann Arbor, MI 48109, USA}
\author[0000-0001-6301-570X]{Benjamin D.\ Boizelle}
\email{boizellb@byu.edu}
\affiliation{Department of Physics and Astronomy, N284 ESC, Brigham Young University, Provo, UT, 84602, USA}
\author[0009-0006-5852-4538]{Hai N.\ Ngo}
\affiliation{Department of Astronomy, University of Michigan, 1085 South University Avenue, Ann Arbor, MI 48109, USA}
\email{hai10hoalk@gmail.com}
\author[0000-0001-5802-6041]{Elena Gallo}
\email{egallo@umich.edu}
\affiliation{Department of Astronomy, University of Michigan, 1085 South University Avenue, Ann Arbor, MI 48109, USA}
\author[0009-0009-0015-1208]{Tuan N.\ Le}
\email{tuan.le.nutshell@gmail.com}
\affiliation{Faculty of Physics—Engineering Physics, University of Science, VNU-HCM, Ho Chi Minh City, Vietnam}
\affiliation{Viet Nam National University Ho Chi Minh City, Vietnam}
\author[0000-0003-1820-2041]{Sabine Thater}
\email{sabine.thater@univie.ac.at}
\affiliation{Department of Astrophysics, University of Vienna, T\"urkenschanzstrasse 17, 1180 Vienna, Austria}
\author[0009-0005-8845-9725]{Tien H.\ T.\ Ho}
\email{htien2808@gmail.com}
\affiliation{Faculty of Physics—Engineering Physics, University of Science, VNU-HCM, Ho Chi Minh City, Vietnam}
\affiliation{Viet Nam National University Ho Chi Minh City, Vietnam}
\author[0009-0004-3689-8577]{Tinh Q.\ T.\ Le}
\email{lethongquoctinh01@gmail.com}
\affiliation{Department of Physics, International University, VNU-HCM, Ho Chi Minh City, Vietnam}
\affiliation{Viet Nam National University Ho Chi Minh City, Vietnam}
\author[0000-0001-9649-2449]{Que T.\ Le}
\email{ltque@hcmiu.edu.vn}
\affiliation{Department of Physics, International University, VNU-HCM, Ho Chi Minh City, Vietnam}
\affiliation{Viet Nam National University Ho Chi Minh City, Vietnam}
\author[0009-0001-2512-1429]{Sam Norcross}
\email{san307@byu.edu}
\affiliation{Department of Physics and Astronomy, N284 ESC, Brigham Young University, Provo, UT, 84602, USA}
\author[0009-0008-1795-3576]{Xueyi Li}
\email{xlilev@udel.edu}
\affiliation{Department of Physics and Astronomy, University of Delaware, 104 The Green Newark, DE 19716, USA}
\author[0009-0008-6050-5736]{Huy G.\ Tong}
\email{tonggiahuy191203@gmail.com}
\affiliation{Faculty of Physics—Engineering Physics, University of Science, VNU-HCM, Ho Chi Minh City, Vietnam}
\affiliation{Viet Nam National University Ho Chi Minh City, Vietnam}
\author[0009-0009-4259-1535]{Nghi K.\ N.\ Le}
\email{lengockhanhnghi@gmail.com}
\affiliation{Faculty of Physics—Engineering Physics, University of Science, VNU-HCM, Ho Chi Minh City, Vietnam}
\affiliation{Viet Nam National University Ho Chi Minh City, Vietnam}
\author[0009-0005-1314-9992]{Huy M.\ B.\ Tran}
\email{tran.ba@ufl.edu}
\affiliation{Department of Astronomy, University of Florida, P.O. Box 112055, Gainesville, FL32611, USA}

\setstcolor{red}

\begin{abstract}

We present ALMA Cycle~7 \cotwo\ observations of the circumnuclear disk in NGC~315 at an angular resolution of $0\farcs230\times0\farcs175$, improving on past measurements and resolving the sphere of influence (SOI) of the supermassive black hole (SMBH), whose mass has previously been estimated of \Mbh~$= \left(2.08^{+0.33}_{-0.15}\right) \times 10^9$ \Msun. The high spatial resolution and sensitivity enable robust full-cube forward modeling of the molecular gas kinematics and a direct comparison of multiple independent gas-based dynamical modeling techniques. We apply standard Bayesian codes using both MCMC and nested sampling approaches, as well as a frequentist code to the same dataset, exploring systematic uncertainties associated with the stellar mass distribution, gas surface-brightness parameterization, and disk geometry. All methods yield consistent black hole masses, indicating that the inferred \Mbh\ is not strongly method-dependent. Combining the ensemble of independent molecular-gas-based models, we derive an ensemble median black hole mass of $M_{\rm BH}/10^9\,\mathrm{M_\odot} = 2.02^{+0.04}_{-0.05}$(stat)$^{+0.05}_{-0.04}$(sys), where the comparable contributions to the full error budget arise from modeling systematics rather than formal fitting uncertainties. Our \Mbh\ is consistent with the empirical \Mbh--$\sigma_\star$ and \Mbh--$L_{\rm bulge}$ scaling relations, and lies 32\% below an independent stellar-dynamical measurement, a discrepancy we discuss in the context of systematic differences between gas- and stellar-based methods. NGC~315 serves as a benchmark for quantifying molecular gas-dynamical \Mbh\ systematic uncertainties and for future cross-comparisons of gaseous and stellar dynamical approaches.

\end{abstract}


\section{Introduction}\label{intro}

Supermassive black holes (SMBHs, with masses \Mbh\;$\gtrsim10^6$ \Msun) reside in the centers of the most massive galaxies (stellar mass $M_\star\gtrsim10^{10}$ \Msun) and seem to co-evolve with their host galaxies \citep[e.g., reviews by][]{Kormendy13, Saglia16, Greene20}. The exact path of this evolutionary history remains enigmatic and may differ across galaxy types \citep[e.g.,][]{Cappellari16, Krajnovic18a}. While scaling relations between SMBHs and properties of their host galaxies have a relatively low scatter, there is an indication that different galaxy types (e.g., ellipticals, spirals, barred galaxies) follow different scaling relations \citep[e.g.,][]{McConnell13, Hartmann2024, Sahu2019}. SMBHs grow mainly either by accretion of gas in its vicinity \citep[e.g.,][]{Fabian12} or the merger of multiple SMBHs \citep[e.g.,][]{Volonteri10}. On the other hand, the main processes that regulate the growth of galaxies are related to in-situ star formation or ex-situ processes, like the accretion of stars or galaxy merging. Unfortunately, frequently large \Mbh\ uncertainties and the limited SMBH sample with well-constrained masses \citep[e.g.,][]{Greene20} prevents a clearer picture of which processes dominate the co-evolution. Determining the masses of SMBHs is a time-intensive task that requires high-quality, high-resolution photometric and spectroscopic observations, often requiring adaptive optics \citep[AO; e.g.,][]{Cappellari2009, Thater17, Thater19, Thater2022, Thater2023, Nguyen14, Nguyen18, Nguyen19, Ngo2025b, Ngo2025c} or interferometric data sets and large computing power \citep{Cappellari2025}. 

To date, approximately 200 black hole masses have been measured dynamically using three classes of methods, each with distinct strengths and limitations. Megamaser accretion disk dynamics \citep[e.g.,][]{Miyoshi95} provide the most geometrically precise measurements, as very long baseline interferometry (VLBI) can resolve Keplerian rotation on sub-parsec scales within the SMBH sphere of influence (SOI), but this method is limited to rare, suitably oriented edge-on disk systems. Stellar dynamics \citep[e.g.,][]{Ahn18, Nguyen17, Nguyen2023, Nguyen_2025_NSC, Nguyen_2026_4258, Nguyen_2026_highz, LeTinh2026} are broadly applicable across galaxy types and mass regimes, but are subject to the mass--anisotropy degeneracy and sensitivity to dust obscuration near the nucleus. Gas-dynamical methods span a wide range in reliability depending on the tracer used. Dynamically warm molecular and ionized gas \citep[e.g.,][]{Verolme02, Seth10, denBrok15} are observationally efficient but susceptible to non-circular motions and turbulence that can bias the inferred \Mbh. Cold molecular gas \citep[e.g.,][]{Davis20, Nguyen20, Ngo2025a, Nguyen_2026_4061} largely avoids these limitations, as it typically resides in geometrically thin, dynamically cold disks that follow nearly circular Keplerian orbits; when the SOI is spatially resolved, this method can achieve a precision comparable to megamaser dynamics \citep[e.g.,][]{Zhang2024}. Atomic gas \citep[e.g.,][]{Nguyen21} offers a complementary approach, though it is less commonly employed. Direct imaging of the event horizon via the Event Horizon Telescope has also provided independent mass constraints for M87 \citep{EventHorizonTelescopeCollaborationM87_1} and Sgr~A* \citep{EventHorizonTelescopeCollaborationMW_1}.

At lower precision, reverberation mapping \citep[e.g.,][]{Barth04, Peterson93} provides \Mbh\ estimates for active galaxies by measuring the light-travel-time delay $\tau$ between continuum variations and the lagged response of broad emission lines, yielding the BLR radius $R_{\rm BLR} = c\tau$. The black hole mass follows as \Mbh~$= f\, R_{\rm BLR}\,\Delta V^2 / G$, where $\Delta V$ is the emission-line width and $f$ is a virial factor encoding the unknown broad line region  geometry, calibrated empirically against local galaxies with direct dynamical \Mbh\ measurements \citep[e.g.,][]{Onken04, Grier13, Batiste17}. Reverberation mapping further anchors the radius--luminosity relation \citep{Bentz13} underlying single-epoch virial estimators, which have been applied to samples of $\gtrsim$$10^5$ AGN in large spectroscopic surveys \citep[e.g.,][]{Shen11}.

The precision and accuracy of all these measurements vary substantially with data quality and modeling assumptions; cross-comparison between independent dynamical approaches is therefore essential for quantifying systematic uncertainties in \Mbh\ determinations. In this work, we focus on \cotwo\ molecular gas kinematics in NGC~315 as a tracer for the black hole's gravitational sphere of influence. The giant early-type galaxy (ETG) NGC~315 is the brightest group/cluster galaxy of the Zwicky cluster \citep{Zwicky61}, which is located at a distance of 72.3 Mpc within the Perseus-Pisces filament. General properties of NGC~315 are summarized in Table \ref{target}. Using the stellar velocity dispersion of $\sigma_\star = 350$~\kms\ for NGC~315 \citep{Pilawa2025}, we apply the $M_{\rm BH}$--$\sigma_\star$ relation of \citet{Kormendy13} for classical bulges and ellipticals, which predicts a SMBH mass of $M_{\rm BH} = 2.9 \times 10^9$~\Msun.

Using Atacama Large Millimeter/submillimeter Array (ALMA) \cotwo\ observations at an angular resolution of $\approx$0\farcs 3, \citet{Boizelle21} measured a mass \Mbh~$= \left(2.08^{+0.33}_{-0.15}\right)\times10^9$~\Msun\ (including both statistical and systematic uncertainties). Based on the standard SOI definition \citep{Merritt2013} where the enclosed total mass is approximately twice the SMBH mass, i.e., $M(<r)=2\times M_{\rm BH}$, the NGC~315 SOI has a radius of $0\farcs67-0\farcs8$ (230--270 pc). This is larger than the approximate $r_g \equiv GM_\mathrm{BH}/\sigma^2_\star \approx 0\farcs2$. In ALMA Cycle~7, we obtained higher-resolution $\approx$$0\farcs2$ observations of the same molecular transition, enabling us to probe down to just $\sim$21~pc scales. These new data resolve the full SMBH sphere of influence by a factor of $\sim$7.6. The combination of improved angular resolution and substantial SOI coverage provides an ideal testbed for assessing systematic uncertainties among different gas-dynamical modelling approaches.

Molecular gas-based \Mbh\  measurements have mainly been obtained with either a Bayesian approach using \kinms\ \citep{Davis13, Davis14} with Markov chain Monte Carlo (MCMC) simulations \citep[e.g.,][]{Davis17, Onishi2017, North19, Smith19, Smith21} or \dynesty\ modelling with nested sampling  \citep{Cohn21, Cohn2024} or frequentist optimization \citep{Barth16a, Boizelle19, Boizelle21}. These different methods have not yet been directly compared in a fully consistent manner. In this study, we aim to address two goals: (i) to obtain a new, more precise SMBH mass measurement of NGC~315 using our recent ALMA observations with a spatial resolution of $0\farcs2$, and (ii) to directly compare the three most commonly applied dynamical modelling methods for molecular gas-based \Mbh\ measurements. This work will serve as a benchmark for future molecular gas-based \Mbh\ measurements.

\begin{table}
  \centering \caption{Properties of NGC~315}
  \vspace{-0.3cm}
    \begin{tabular}{lcr}
    \hline\hline
    Parameter (units) & Value & References\\
    \hline
    Morphology    & cD: & (1) \\
    R.A.\ (J2000) &  00$^{\rm h}$57$^{\rm m}$48.8834$^{\rm s}$ &(2)\\
    Decl. (J2000) & +30\degr21\arcmin08\farcs8119 & (2)\\
    $cz_{\rm obs}$ (\kms) & 4942.1 & (3) \\
    Stellar Isophotal PA (\degr) & 44.3\degr & (4)\\
    Stellar Ellipticity & 0.27 & (4) \\
    Luminosity Distance (Mpc)  & 72.3 & (5) \\
    Linear Scale (pc arcsec$^{-1}$)  & 340 &  (5) \\
     \hline
  \end{tabular}
  \parbox[t]{\columnwidth}{\footnotesize {\bf Notes:} (1) \citet{deVaucouleurs91}; (2) \citet{Kovalev05}; (3) \citet{Trager00}; (4) luminosity-weighted values from \citet{Goullaud18}; (5) \citet{Boizelle21}.}
  \label{target}
\end{table}

Several galaxies now have \Mbh\ measurements from both stellar and molecular gas dynamics, enabling direct cross-method comparisons. These include NGC~404 \citep{Nguyen17, Davis20}, NGC~524 \citep{Krajnovic09, Smith19}, NGC~1332 \citep{Rusli11, Barth16}, NGC~4697 \citep{Schulze11, Davis17}, NGC~4751 \citep{Rusli13, Dominiak2025}, and NGC~6861 \citep{Rusli13, Kabasares22}. In most cases, the two methods yield consistent \Mbh\ values within their respective uncertainties, providing mutual validation of both approaches and demonstrating that well-resolved molecular gas dynamics can serve as a reliable cross-check against stellar dynamical modeling.

Recent stellar-dynamical modelling of NGC~315 has provided an independent measurement of its SMBH mass. Using high signal-to-noise integral-field spectroscopy obtained at the Gemini and McDonald Observatories, \citet{Pilawa2025} measured spatially resolved stellar kinematics (PSF FWHM of 0\farcs3) across the inner $\sim$30\arcsec\ of the galaxy. They constructed triaxial Schwarzschild orbit-based models with sped-up Latin hypercube sampling to find a strongly prolate intrinsic shape and \Mbh~= $(3.18\pm0.32)\times10^9$~\Msun, which is discrepant with previous molecular gas-based measurements by $\sim1.7\sigma$ \citep{Boizelle21}. This target is also the focus of even higher-resolution ALMA gaseous (project 2022.1.01040.S; PI: Boizelle) and James Webb Space Telescope (JWST) stellar-dynamical studies (GO-5716; PI: Walsh). The availability of both stellar- and gas-dynamical constraints makes NGC~315 an especially valuable system for assessing the consistency and potential systematics of different SMBH mass measurement techniques.

This paper is organized as follows. In Section \ref{observations}, we introduce our new ALMA observations from Cycle~7. These observations were extensively modelled with different dynamical modelling methods as explained in Section \ref{s:modelling}. We discuss our results and the implications in Section \ref{discussion}, and summarize the main results in Section \ref{conclusion}. We adopt a luminosity distance $D_L = 72.3$ Mpc for NGC 315. For a redshift $z_\mathrm{obs} \sim 0.0161578$ that is corrected for the Virgo + Great Attractor + Shapley Supercluster inflow model \citep{mould00}, a standard cosmology \citep{Planck16} results in an angular scale of 340 pc arcsec$^{-1}$ \citep{Boizelle21}.

\section{Data}\label{observations}

\subsection{Photometric Observations $\&$ Stellar Luminosity Profile}\label{mge}

\begin{figure}
  \centering\includegraphics[scale=0.6]{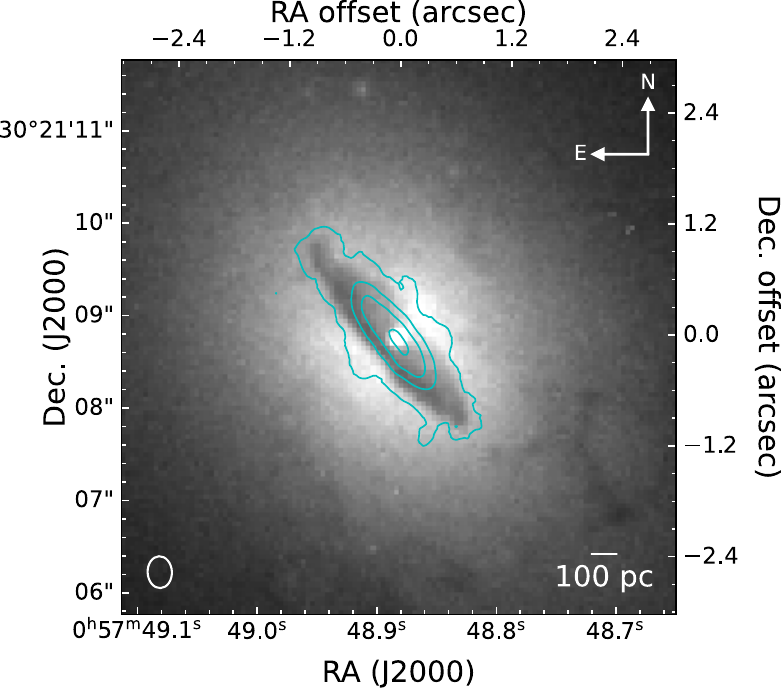}
  \caption{\hst/WFC3 F110W imaging of NGC~315 shown together with the ALMA \cotwo\ moment 0 surface brightnesses (cyan contours). The CND is well traced by both CO emission and dense dust absorption, with some faint, filamentary features the south-west direction only detected by dust.}
  \label{hst-alma}   
\end{figure}

To apply state-of-the-art gas-dynamical modeling to NGC~315 circumnuclear disk (CND) traced by \cotwo, we adopted stellar luminosity models constructed by \citet{Boizelle21}. The \emph{Hubble Space Telescope} ({\it HST}) Wide Field Camera 3 \citep[WFC3;][]{Dressel19} F110W mosaic \citep[in GO--14219;][see Figure \ref{hst-alma}]{Goullaud18} is supplemented at large radii with scaled {\it Spitzer} InfraRed Array Camera \citep[IRAC, 3.6 $\mu$m channel;][]{Fazio04} data from the {\it Spitzer} Enhanced Imaging Products \citep{irsa433}. While these NIR filter passbands do not overlap, they trace the old stellar population with minimal color gradients in the stellar halo. Wide Field Planetary Camera 2 \citep[WFPC2;][]{Holtzman95} optical imaging in both F555W and F814W bands \citep[in GO--6673;][]{VerdoesKleijn99} were also used to identify and correct for the range of plausible of intrinsic dust obscuration (central $A_{\rm F110W} = 0,\,0.75\,1.50$ mag, corresponding to a 0, 25, and 33\% loss, respectively) of stellar light that originated \textit{behind} the disk. A dust-masked image and those that were dust-corrected were then decomposed into stellar luminosity models using the Multi-Gaussian Expansion \citep[MGE;][]{Emsellem94, Cappellari02} formalism in \galfit\ \citep{Peng10} while also incorporating a point source to account for active galactic nuclei (AGN) contributions. We refer the reader to Section~2 and Table~1 of \citet{Boizelle21} for additional details concerning the calibration processes, surface brightness measurements, dust corrections, and MGE results.

\subsection{ALMA Observations}\label{almaobs}

\subsubsection{Calibrations}\label{calibration}

NGC~315 has \cotwo\ observations from three ALMA projects were designed to resolve and characterize the rotating CO-bright disk properties. \citet{Boizelle21} detailed Cycle~5 \cotwo\ imaging (PID: 2017.1.00301.S; PI: Barth) obtained with a geometrically-averaged beam full width at half maximum (FWHM) $\overline{\theta}_\mathrm{FWHM} \approx 0.3\arcsec$ ($\sim$102 pc). Standard gas-dynamical modelling returned a best-fitting \Mbh~$= \left(2.08^{+0.33}_{-0.15}\right) \times 10^9$ \Msun, including major systematic uncertainties. In this paper, we present Cycle 7 \cotwo\ imaging of the same CND (PID: 2019.1.00036.S; PI: Dieu Nguyen) that was carried out on 2021 July 01. This scheduling block spanned a total time of 55 min, with 29 min spent on-source in a single spectral setup, including latencies. The correlator was set up using four frequency division mode (FDM; 1875 MHz bandwidth with 1.13 MHz channels) spectral windows (spw). One spw was centered on the \cotwo\ line while the other three were placed to probe continuum emission across the 223.86--241.97 GHz range. A total of 44 12-m array antennae in the C43--6 configuration resulted in baselines ranging from 15--2280 meters. Flux and bandpass calibrations were carried out using the quasar J0238+1636, while atmospheric phase offsets were determined using the quasar J0048+3157. The raw ALMA data were calibrated using the standard pipeline using version 6.1.1--15 of the \texttt{Common Astronomy Software Applications} \citep[\casa\footnote{\url{https://casadocs.readthedocs.io/en/stable/notebooks/citing-casa.html}};][]{McMullin07}. After the pipeline procedures, we applied iterative continuum phase and amplitude self-calibration processes with final \texttt{gaincal} solution intervals being equal to twice the integration length and the scan length, respectively. After self-calibration, we modeled and removed the continuum in the \textit{uv} plane with the \texttt{uvcontsub} task using line-free channels. On all image-plane products, we applied primary beam corrections.

\begin{figure*}[!th]
  \centering\includegraphics[width=0.98\linewidth]{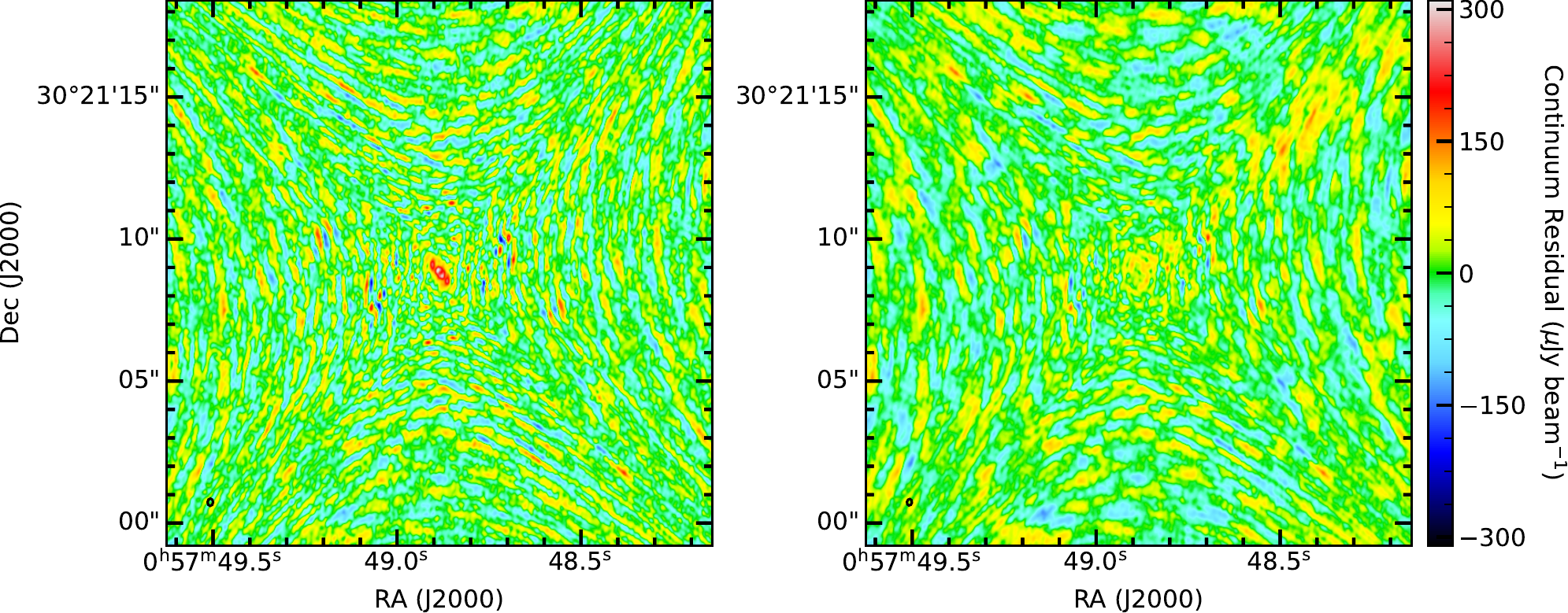}
  \caption{Continuum residual maps of NGC~315 after subtracting both the central point source and the extended thermal-dust component (modelled as a Gaussian; PA~$\approx35^\circ$), shown for the ALMA Cycle~5 data (\textit{left panel}) and the Cycle~7 data (\textit{right panel}). Both maps use natural weighting and the same point-source-plus-dust subtraction procedure (Section~\ref{cont}), so the comparison isolates the improvement afforded by the higher-resolution Cycle~7 data: the Cycle~7 residual (RMS $\sim$35~$\mu$Jy~beam$^{-1}$) shows a substantially cleaner subtraction at the nucleus than the Cycle~5 map. In the Cycle~7 residual, a faint feature extending at PA~$\approx-45^\circ$ may trace the approaching radio jet.}
  \label{cont_residuals}   
\end{figure*}

\subsubsection{Band 6 Continuum Emission}\label{cont}

\begin{figure*}[!th]
  \centering\includegraphics[scale=0.8]{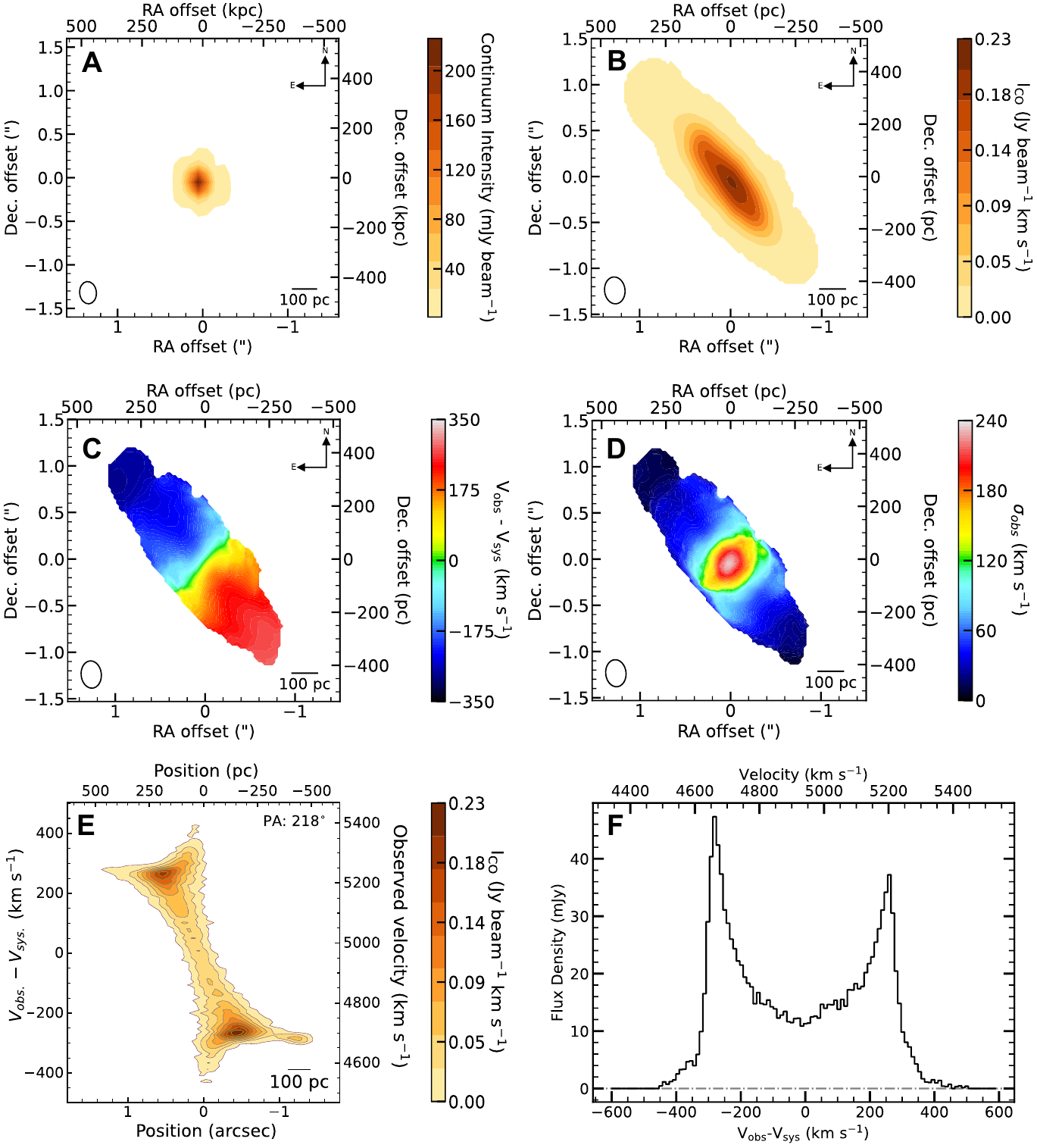}
  \caption{ALMA Band 6 continuum and \cotwo\ imaging of NGC~315 with a synthesized beam $\theta_{\rm beam}=0\farcs230\times0\farcs175$ (or $\approx$$78\times60$ pc$^2$) is shown as a black ellipse in the bottom-left corner of each panel. {\it Panel A} shows the centrally-concentrated 1.3 mm continuum emission. In contrast, others show the integrated intensity (moment 0, {\it Panel B}), intensity-weighted mean line-of-sight (LOS) velocity (moment 1, {\it Panel C}), and intensity-weighted LOS velocity dispersion (moment 2, {\it Panel D}) maps of the \cotwo\ molecular gas emission in NGC~315. All emission detected within the region of $r\approx1\farcs5$ (or $\approx$510~pc). These maps are centred on (R.A., Decl.)$\,=(0^{\rm h}57^{\rm m}48\fs883$, $+30\degr21\arcmin08\farcs812$). {\it Panel E} shows the PVD of the \cotwo\ emission extracted along the major axis with a slit width of ten pixels ($0\farcs3$ or 102 pc). {\it Panel F} shows the integrated CO spectrum extracted within the nuclear region of $1\farcs5\times1\farcs5$ (or 510 $\times$ 510 pc$^2$) region, which reveals the classical symmetric double-horn shape of a rotating disk.}
  \label{comaps}   
\end{figure*}

Band 6 continuum emission ($\approx$1.3 mm) is clearly detected at the centre of NGC~315 with a hint of extended continuum in the disk direction (Panel~A, Figure~\ref{comaps}). Multi-frequency synthesis (MFS) mode imaging over the full line-free bandwidth results in a synthesized beam with FWHM $\theta_\mathrm{maj}\times \theta_\mathrm{min} = 0\farcs 230 \times 0\farcs 175$ at a PA of 2.7\degr. Using \texttt{tclean} deconvolution \citep[Briggs weighting with $r = 0.5$;][]{Briggs95}, the peak flux density of $\sim$226 mJy measured over the full line-free bandwidth is slightly higher than the value measured towards the end of 2018 \citep[although they remain consistent within absolute calibration uncertainties;][]{Boizelle21}. Interactive masking and \texttt{CLEAN}ing down to $\sim$4$\times$ the root-mean-square (rms) level help to reduce sidelobe noise. The final limiting sensitivity of $\sim$43 $\mu$Jy beam$^{-1}$ is slightly improved compared to the Cycle~5 continuum emission, due in part to a much lower precipitable water vapor (PWV) column of 0.3 mm. The final MFS image peak-to-rms reaches a dynamic range of nearly 5000.

We fit this continuum-only image with two elliptical Gaussian components using the {\tt CASA} {\tt imfit} task to model both the prominent central source as well as continuum emission arising from the resolved disk. The central source is found to have a centroid of (R.A., Decl.)$\,=(0^{\rm h}57^{\rm m}48\fs883$, $+30\degr21\arcmin08\farcs812$) that is consistent with both peak radio-to-mm continuum locations from VLBI imaging \citep[e.g.,][]{Boccardi21} and the CO disk kinematic center \citep{Boizelle21}. Its fitted major and minor-axis FWHM are $0\farcs 230 \times 0\farcs 175$ and the PA is $\sim$2.6\degr, which closely matches the synthesized beam properties. The extended elliptical Gaussian component has a consistent centroid but much larger FWHM of $0\farcs974\times 0\farcs371$ orientated at a PA of 34.4\degr\ that agrees with the disk orientation and extent. Its integrated flux density of 1.96 mJy is likely thermal emission from cold dust grains within the CND.

Following the method outlined by \citet{Boizelle17}, we fit and subtracted first the central continuum point source and then also the disk-like component using \texttt{uvmodelfit}, with the model point source flux adjusted to avoid oversubtraction due to residual thermal emission from the circumnuclear disk.   We applied this same subtraction procedure to both the Cycle~5 and Cycle~7 datasets; the resulting residual maps are shown in the left and right panels of Figure~\ref{cont_residuals}, respectively. The Cycle 7 residual map (natural weighting; RMS $\sim$35~$\mu$Jy~beam$^{-1}$; right panel of Figure~\ref{cont_residuals}) shows a substantially cleaner subtraction at the nucleus than the Cycle 5 map, and may reveal faint extended continuum emission stretching from $\sim$0\farcs75 to $\sim$9\arcsec\ from the nucleus at PA $\approx -45\degr$, consistent with the orientation of the approaching radio jet \citep{Morganti09, Boccardi21}. While only slightly above the noise level, its asymmetry relative to the other three corners of the image lends tentative support to a jet-related rather than instrumental origin.

\begin{figure}
  \centering
  \includegraphics[scale=0.35]{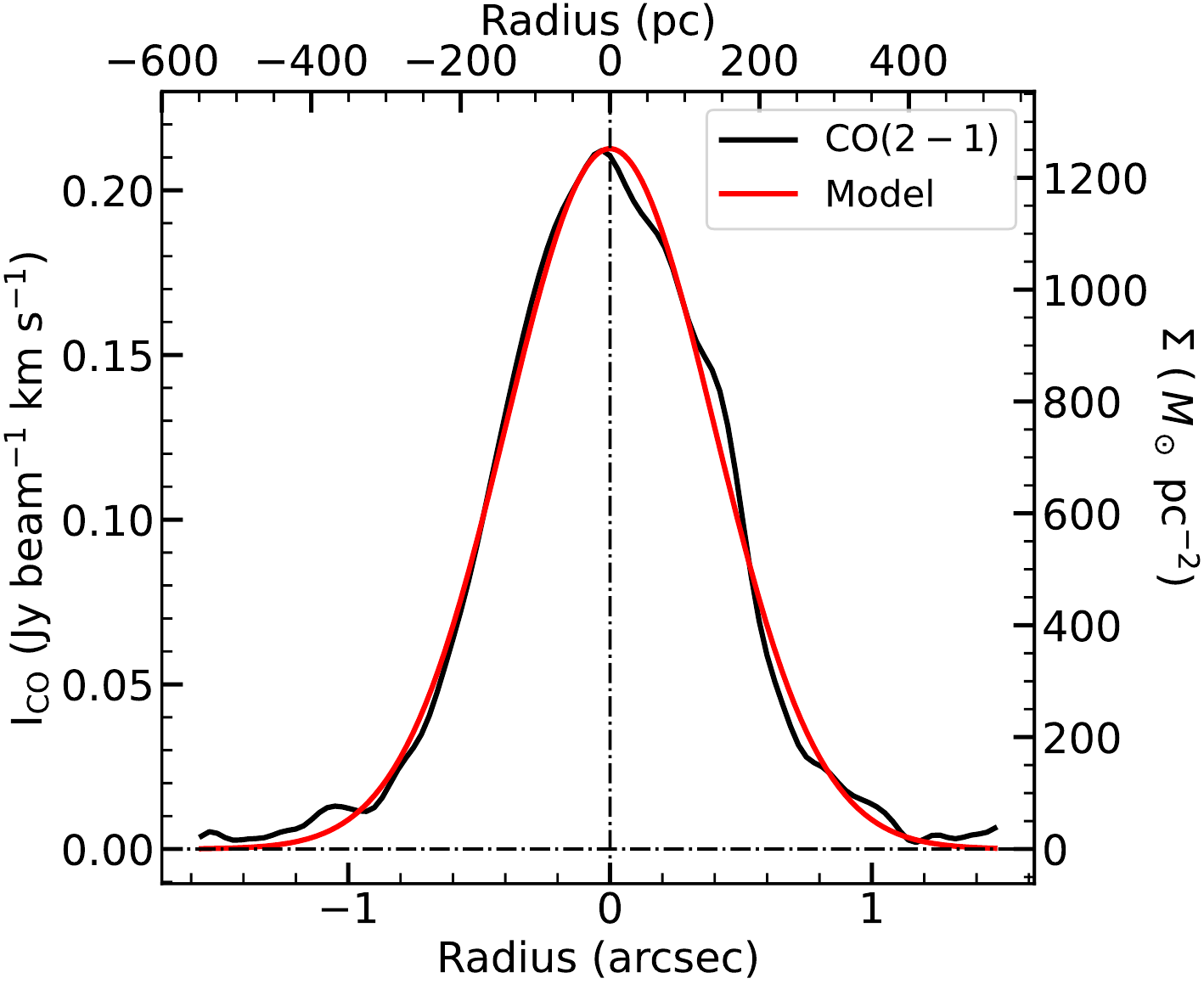}
   \caption{The \cotwo\ morphological distribution in NGC~315 is fairly symmetrical about the continuum center at $\sim$102 pc resolution, as is shown along the major axis of the integrated intensity map (Panel B of Figure~\ref{comaps}). A Gaussian surface brightness model matches the distribution well.}
  \label{GaussianSB}    
\end{figure}

\subsubsection{\cotwo\ Moment Maps}\label{momentmaps}

We deconvolved the continuum-subtracted calibrated measurement set using an iterative \texttt{tclean} process and the same Briggs weighting to create a three-dimensional (3D; R.A., Decl., velocity) ALMA \cotwo\ datacube. We selected a pixel size of $0\farcs03$ based on the beam $\theta_\mathrm{maj} \times \theta_\mathrm{min} = 0\farcs 230 \times 0\farcs 175$ at a PA of 6.4\degr, which is a factor of about $\approx$1.5$\times$ smaller than the Cycle 5 cube beam FWHM. The binned channel width of 10 km $^{-1}$ in the observed frame was intended to appropriately sample the observed lower CO line FWHM \citep[$\sim$30 \kms;][]{Boizelle21}. Our final \cotwo\ datacube achieved an rms noise of about 0.36 mJy beam$^{-1}$ per 10 \kms\ channel, essentially the same as the Cycle~5 sensitivity. We note that tests combining the Cycles~5 and 7 imaging of NGC~315 into a single \cotwo\ datacube degrade the angular resolution while only slightly improving the limiting sensitivity. As this project serves as an independent test of the fidelity of CO-derived \Mbh, we restrict the gas-dynamical analysis and modelling to the Cycle 7 data only.   A dedicated study of the higher-resolution ALMA data (2022.1.01040.S; PI: Boizelle) will be presented separately. Combining datasets of differing angular resolution would conflate variations in data quality with those in modelling methodology, obscuring the modelling systematics that are the focus of this work. Precisely because the Cycle~7 data resolve the SOI well but not to the degree the higher-resolution data will, the modelling systematics, in particular the degeneracy between the SMBH and the central stellar mass distribution, remain non-negligible. This makes the dataset well suited to a method-comparison study: a measurement in which the SOI is resolved so completely that all methods trivially agree would provide little leverage for isolating the modelling systematics we set out to quantify.

The \cotwo\ emission is clearly detected in channels spanning about 4460--5450 \kms\ with an average velocity of $\approx$4945 \kms. Figure~\ref{comaps} shows the \cotwo\ integrated intensity (moment 0), intensity-weighted line-of-sight (LOS) velocity (moment 1), and intensity-weighted LOS velocity dispersion (moment 2) maps of the CND region. We created these maps directly from the 3D \cotwo\ cube using the moment-masking technique \citep{Dame01, Dame11} with detailed implementation clearly discussed by \citet{Nguyen21, Nguyen22}. In general, these moment maps are mostly consistent with those from the Cycle~5 data. We constructed a position--velocity diagram (PVD) by extracting a slit of 10 pixels in width ($0\farcs3$ or 100~pc) through the CND's kinematic major axis (at a line-of-nodes PA $\Gamma=218^\circ$) along with a spectrum (Panel F) integrated over the central $1\farcs5\times1\farcs5$ (or 510 $\times$ 510 pc$^2$) region that shows a classical double-horn shape of a rotating disk.

The integrated intensity map shows smoothly varying, centrally concentrated \cotwo\ emission, with a total flux of $11.70\pm 1.17$ Jy \kms\ \citep[with total uncertainty dominated by the absolute flux calibration uncertainty;][]{Francis20} that is consistent with the Cycle 5 measurement \citep{Boizelle21}. As is shown in Figure~\ref{GaussianSB}, the \cotwo\ fluxes extracted along the major axis of the moment 0 map is well fitted by a Gaussian function. The intensity-weighted LOS velocity map shows a rotating disc with minor kinematic twists. While the luminosity-weighted LOS velocities only span a total $\Delta V\approx700$ km s$^{-1}$, \cotwo\ emission is detected out to about $\pm$435 \kms. Compared to \cotwo\ from Cycle 5 data, the Keplerian wings in the Cycle~7 PVD are detected up to $\sim$20 \kms\ further out. The intensity-weighted LOS velocity dispersion map is centrally concentrated, increasing from a FWHM~$\sim 30$ \kms\ up to $\sim$560 \kms\ due to beam smearing and strong rotational broadening. Thus, except for the slight increase in velocity range, the structure of the moment maps from our data is fully consistent with those found by \citet{Boizelle21} using lower spatial resolution~data. 

The integrated intensity map shows that the CND extends $\approx$3\arcsec\ along the major axis and 1\arcsec\ along the minor axis, featuring a smooth intensity gradient and peaks at the CND's center. It also illustrates the alignment between the \cotwo\ emission and the dust disk (see Figure~\ref{hst-alma}).  The total molecular gas mass was estimated by using the CO-to-H$_2$ conversion factor of $X_{\rm CO(1-0)} = 2 \times 10^{20}{\,\rm cm^{-2}\,(K\,km~s^{-1})^{-1}}$ \citep[or $\alpha_{\rm CO(1-0)} = 4.3$ \Msun\ (K\,\kms\ pc$^{-1}$)$^{-1}$;][]{Bolatto2013}:
\begin{equation} 
\small
	\begin{split} 
M_{\rm gas}/M_\odot &= 1.05 \times 10^4\left( \frac{X_{\rm CO(1-0)}}{2 \times 10^{20} \dfrac{\rm cm^{-2}}{\rm K\ km\ s^{-1}}} \right) \\
& \times\left( \frac{1}{1+z} \right) \left( \frac{S_{\rm CO(1-0)}\Delta v}{\rm Jy\ km\ s^{-1}} \right) \left( \frac{D_L}{\rm Mpc} \right)^2 .
	\end{split} 
\end{equation}
This requires conversion of the integrated \cotwo\ flux to the expected CO(1$-$0) value $S_{\rm CO(1-0)}\Delta v \approx 2.93$ Jy \kms\ assuming a \cotwo\ and \coone\ flux ratio of about unity \citep[in K units;][]{Sandstrom13, Ngo2025a, Nguyen_2026_4061}. For this calculation, we adopted $z \approx 0.016485$ for NGC 315 from the NASA/IPAC Extragalactic Database\footnote{NED: \url{https://ned.ipac.caltech.edu}}), which matches a luminosity distance of $D_L = 72.3$~Mpc \citep{Boizelle21} and assume a unity flux ratio between \cotwo\ and \coone\ \citep{Ngo2025a, Nguyen_2026_4061}. We further correct the inferred H$_2$ mass for the helium contribution by adopting a helium mass fraction of $f_{\rm He}=0.36$. Under these assumptions, we derive a total molecular gas mass of $M_{\rm gas} \approx (2.15 \pm 0.18) \times 10^8$~\Msun\ that is consistent with the value reported by \citet{Boizelle21}.

\section{Dynamical Black Hole Mass Measurements}\label{s:modelling}

To measure the NGC~315 SMBH mass, we apply a now standard forward-modelling approach to create a model cube and optimize its parameters by fitting directly to the ALMA Cycle~7 \cotwo\ datacube. This approach simultaneously constrains \Mbh\ and the disc kinematic properties as well as the host galaxy mass profile. In this section, we briefly describe the different modelling codes we employed to test the consistency of measured \Mbh\ values. These codes vary primarily in their use of input flux maps and the treatment of the disc structure as either intrinsically flat or warped. All models incorporate beam-smearing effects before comparing model and data cubes. For additional details of the general flat-disc model approach, we refer the reader to the discussions by \citet{Barth16b, Barth16a} and \citet{Davis17}; for additional details on the treatment of a warped disc using a tilted-ring model, refer to \citet{Boizelle19}. In the following, we find best-fit \Mbh\ values and explore various sources of systematic uncertainty to determine a final SMBH mass error budget for NGC~315.

We assume the gas moves on circular orbits governed by the gravitational potentials of at least the SMBH and the stellar mass distribution. The latter is calculated using the \texttt{mge\_circular\_velocity} task within the Jeans Anisotropic Modelling\footnote{\url{https://purl.org/cappellari/software}} \citep[JAM;][]{Cappellari08} package in the \texttt{Interactive Data Language (IDL)}. This task calculates circular velocities in the galaxy midplane using an axisymmetric mass model specified by a MGE \citep{Cappellari02} after converting surface brightnesses by a stellar mass-to-light ($M/L$) ratio. In this paper, we adopt a constant $M/L$ ratio at all radii and employ MGE solutions constructed previously using \hst\ WFC3/F110W imaging \citep[mentioned in Section \ref{mge};][]{Boizelle21}. In most cases, our dynamical models use the MGE solution that corrects for an intermediate amount of dust ($A_{\mathrm{F110W}} = 0.75$ mag; their model B2); in Section~\ref{additional}, we employ additional dust-masked and dust-corrected MGEs to determine the effect of circumnuclear dust on the best-fit \Mbh\ value.

\subsection{Flat-disc Models}\label{flatdisk}

In Figure~\ref{comaps}, the \cotwo\ moment 1 line of nodes orientation appears essentially constant ($\Delta\Gamma\lesssim 5\degr$) for $R<1\arcsec$ with slightly more noticeable kinematic twists ($\Delta\Gamma \sim 10\degr$) towards the disc edge (see also Figure~\ref{ngc315_almac7m_kintwist}). Results from  \citet{Boizelle19} suggest that, for mildly warped discs, the best-fitting \Mbh\ is not likely to be strongly biased by the adoption of a flat-disc model for the gas distribution. Our first \cotwo\ datacube modelling efforts, therefore, assumed the gas distribution to be arranged in a geometrically flat disc.

This flat-disc approach requires a minimum of $P=13$ free parameters to create the model cube. The flat disc itself is defined spatially by an ($x_\mathrm{c}$,$y_\mathrm{c}$) centroid location together with position angle ($\Gamma$) and inclination angle ($i$). The gas kinematics are defined by the total gravitational potential that is dominated by \Mbh, the stellar $M/L_{\mathrm{F110W}}$ ratio, and the total molecular gas mas $M_{\rm gas}$. The LOS rotational map is then shifted by the galaxy's systemic velocity $\vsys = cz = 4942.1$ \kms\ \citep[from stellar absorption-line fitting;][]{Trager00} and an additional $v_\mathrm{off}$ offset with positive values indicating greater redshift. The model cube Gaussian line profiles are weighted by an input surface brightness map with a total flux $f$ that closely matches the integrated moment 0 value. 

The intrinsic turbulent velocity width $\sigma_{\rm turb}$ represents the local line-of-sight velocity dispersion of the molecular gas and may vary with radius. For this Cycle 7 modelling, we optimized the same Gaussian parametric radial profile adopted by \citet{Boizelle21} to the Cycle~5 data of NGC 315. Their $\sigma_{\rm turb} = \sigma_0\exp[-(r-r_0)^2/2\mu^2] + \sigma_1$ function returned a high-amplitude but extremely centrally concentrated Gaussian component ($\sigma_0 \sim 120$~\kms, $\mu \lesssim 30$~pc $< 0\farcs1$) superimposed on a uniform disk-wide line width of $\sigma_1 \sim 15.2$~\kms.

Here, the model optimization is carried out using either a Bayesian framework \citep[using either the Python \kinms\ version \footnote{\url{https://github.com/TimothyADavis/KinMS_MCMC}} or \dynesty\ codes;][]{Davis13, Davis14, Davis2020ascl.soft, Speagle20} or using simple $\chi^2$ minimization by a downhill simplex method \citep{Press92}, where the goodness-of-fit 
\begin{equation}
    \chi^2 = \sum_{i=1}^N \frac{({\rm data}_i - {\rm model}_i)^2}{\delta^2}
\end{equation}
\noindent is measured over all $i$ pixel locations for $N$ constraints. The noise $\delta$ is estimated from the RMS in line-free regions of each \cotwo\ cube channel and is generally assumed to be a constant value for each frequency slice \citep[however, c.f.,][]{Kabasares22}. The downhill simplex method is used to more quickly explore various systematic uncertainties as well as to optimize the tilted-ring approach described in Section~\ref{tiltedmodel}.

In both Bayesian and $\chi^2$ approaches, we adopt flat priors with all parameters sampled uniformly in linear space except for the SMBH and gas masses that are sampled linearly in $\log_{10}$~\Mbh\ and $\log_{10} M_\mathrm{gas}$ to ensure efficient sampling of the broad prior range. The Bayesian likelihood is determined by $\mathcal{L} \propto \exp(-\chi^2/2)$. Furthermore, the model is optimized by fitting to the datacube directly, with the goodness-of-fit at each model iteration calculated in a 3D fitting region that covers the entire CND in the spatial dimensions and that spans the \cotwo\ emission along the frequency axis. In the \kinms\ and \dynesty\ modelling runs, we use a fixed $3\arcsec\times 3\arcsec$ ($100\times 100$ pixel$^2$) spatial region for each of the $N_{\rm chan}\sim$ 96 channels that spans $\approx\pm 430$ \kms\ from the galaxy's systemic velocity \vsys, resulting in a final number of constraints $N_{\tt KinMS} = 100\times 100\times N_{\rm chan} = 960,000$.

For the $\chi^2$ approach as well as one trial using \dynesty, we follow \citet{Boizelle21} and employ an elliptical spatial fitting region that is tailored to the disc shape, with major-axis $\Gamma = 218\degr$ and major and minor axes extents of $R_{\rm maj} = 1\farcs 59$ and $R_{\rm min} = 0\farcs 41$, respectively. This elliptical region is repeated in each of the same $N_{\rm chan}$ to construct the fitting region. The modelling codes using either \dynesty\ or $\chi^2$ optimization subdivide each spatial pixel into $s\times s$ sub-pixels when constructing the kinematic map and model cube to better capture sub-pixel velocity gradients. After returning the model cube to the native spatial scale, we block-average both the data and model cubes by an integer factor $d$ in both spatial coordinates to mitigate the impact of correlated noise between adjacent pixels prior to calculating the goodness-of-fit. For all but one of the \dynesty\ runs, we chose $s=1$ and $d=1$ to match the native pixel scale and channel width adopted in the \kinms\ modelling. For the $\chi^2$ approach, we used a factor $s = 4$ and $d=4$ to form cells that are closer to the beam size. This results in a smaller number of constraints $N_{\chi^2}=\pi R_{\rm maj}R_{\rm min}/d^2\times N_{\rm chan} = 12,240$ that more closely follow the CO emission. The result will be less dilution of a reduced $\chi^2_{\rm dof} = \chi^2 / (N-P)$ value for $N-P$ degrees of freedom (dof).

\begin{figure*}[!th]
  \centering
  \includegraphics[width=0.9\textwidth]{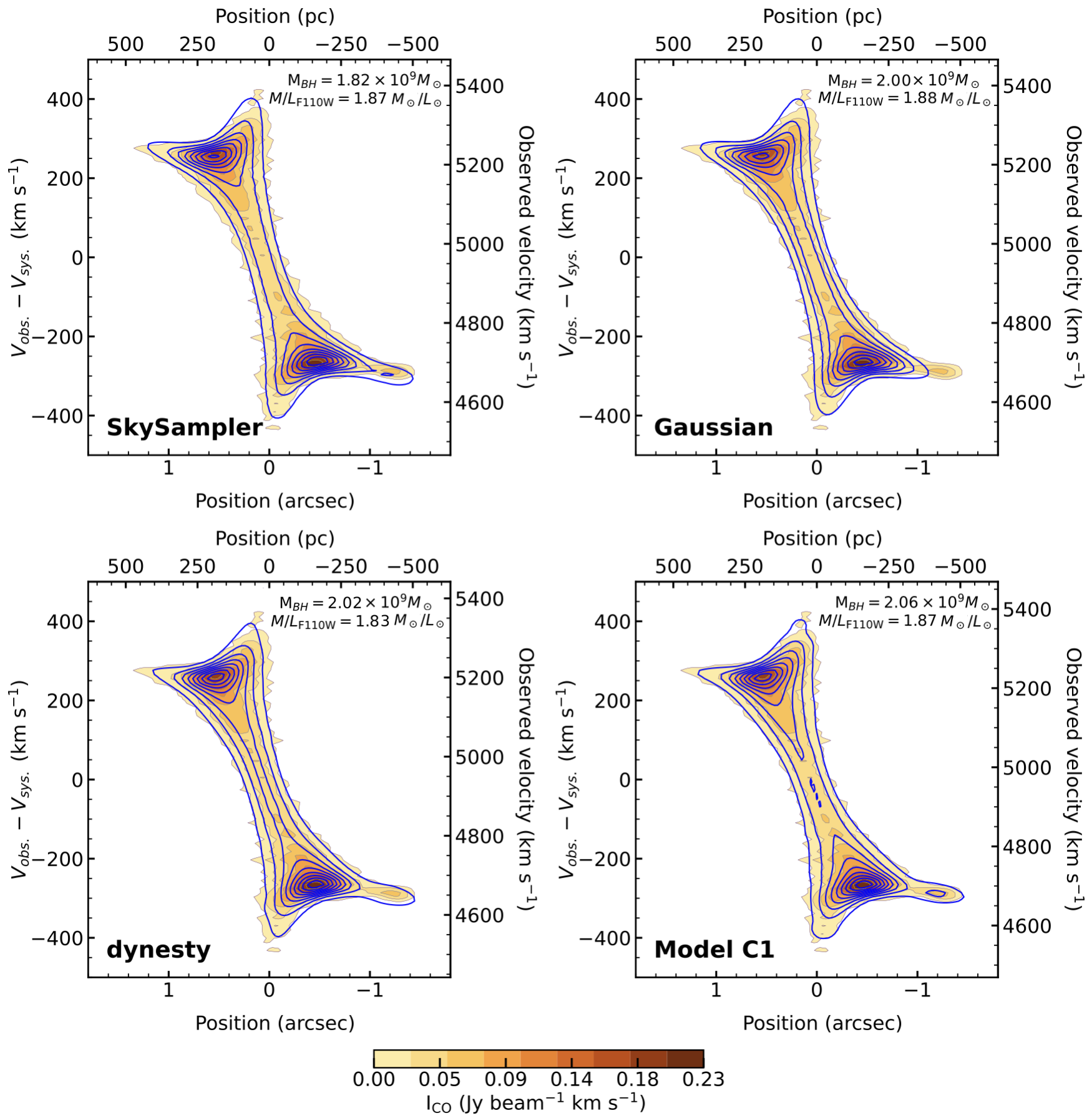}
  \caption{Results of our gas-dynamical modelling of the NGC~315 ALMA \cotwo\ datacube. The \cotwo\ surface brightness distribution is modeled using the \skysampler\ tool within the \kinms\ framework ({\it upper-left panel}), a Gaussian parameterization within \kinms\ ({\it upper-right panel}), the nested-sampling \dynesty\ model ({\it lower-left panel}), and the standard warped-disc model (Model~C; {\it lower-right panel}). All panels present PVDs of the \cotwo\ datacube extracted along the kinematic major axis (orange colorscale), overlaid with the best-fitting model PVDs (blue contours). The corresponding best-fit values of $\ml_{\rm F110W}$ and $M_{\rm BH}$ are indicated in the legend of each panel, while the remaining best-fitting parameters are listed in Table~\ref{kimsbest}. }
  \label{all_best_models}   
\end{figure*}

The \kinms, \dynesty, and $\chi^2$ approaches follow mostly the same general model cube method, including the same input parameters and goodness-of-fit analysis. In Table~\ref{kimsbest}, we report median values of the posterior distribution for each parameter, which are close ($<$1 per cent different in some cases) to the best-fit values determined by the minimum $\chi^2$ location. However, they do differ in the treatment of the frequency or velocity axis. \kinms\ requires the velocity binning and offsets the Doppler-shifted velocity map by an additional \voff\ from the central channel velocity of the \cotwo\ data cube. This treats the observed line widths as if they are rest-frame widths. The \dynesty\ and $\chi^2$ methods construct the model cube on the same frequency grid as the data and transform the rest-frame LOS rotation and line width maps to the observed frame by the variable $cz_\mathrm{obs}$.

When working with very large data sets and high $N$ constraints, the parameter uncertainties are often severely underestimated by a standard statistical error analysis. Instead, we derive statistical uncertainties based on the expected standard deviation of the goodness-of-fit \citep{vandenBosch09}. In the $\chi^2$ approach, we increase the $\Delta\chi^2$ difference required to define a given confidence level (CL) by the $\chi^2$ standard deviation, or $\sqrt{2(N-P)}\approx\sqrt{2N}$ \citep{Press07}. In the Bayesian approach, an equivalent effect can be achieved by dividing the model log likelihood by $\sqrt{2N}$ or, equivalently, multiplying the measurement uncertainties (RMS) by $(2N)^{1/4}$ \citep{Mitzkus17, Nguyen2025a}. This also helps the statistical uncertainties be more similar to the systematic ones. In this paper, we report both 1 and $3\sigma$ (statistical) uncertainties for all parameter values.

\begin{figure*}[!th]
  \centering
   \includegraphics[scale=0.65]{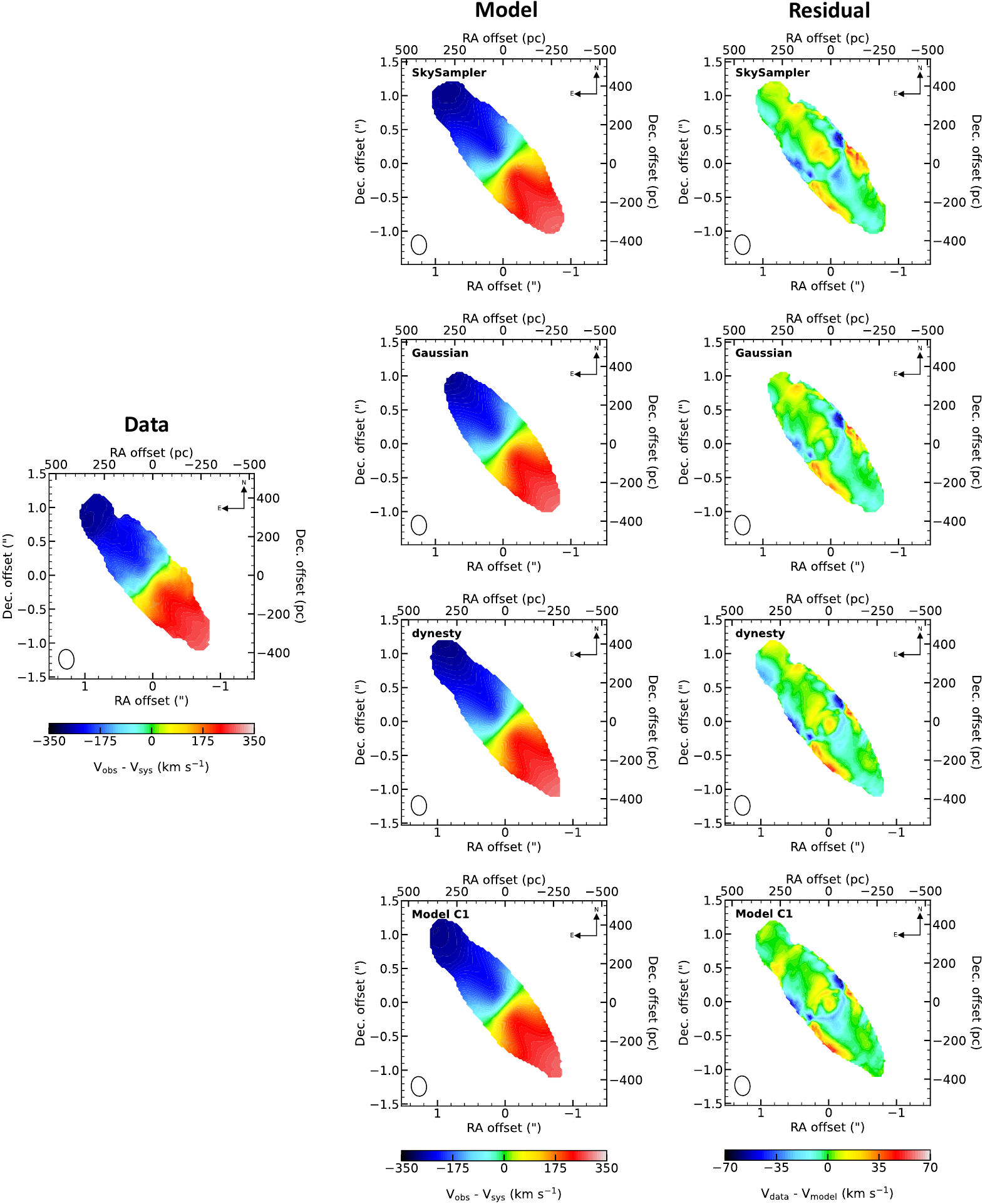}
   \caption{Comparison between the NGC~315 ALMA \cotwo\ datacube and our gas-dynamical models in terms of the mean LOS velocity maps. The observed mean LOS velocity map, identical to that shown in Panel C of Figure~\ref{comaps}, is presented in the {\it left-column panels}. The corresponding best-fitting models are shown in the {\it middle-column panels}, with legends indicating the specific modelling approach adopted. The best-fit values of $\ml_{\mathrm{F110W}}$, $M_{\rm BH}$, and others are consistent with those reported in Figure~\ref{all_best_models} and Table~\ref{kimsbest}. The {\it right-column panels} display the residual maps, defined as {\tt (data$-$model)}, between the observed mean \cotwo\ LOS velocity map and the corresponding best-fitting gas-dynamical models.}   
   \label{all_bestmom1_models_reds}
\end{figure*}

\subsubsection{\kinms\ Modelling}\label{kinms}

We first applied the \kinms\ dynamical modelling code \citep{Davis13, Davis14} to create and optimize a simulated cube to the ALMA \cotwo\ datacube for NGC~315. \kinms\ explores the parameter space efficiently using a Markov chain Monte Carlo (MCMC) method that is controlled by the \texttt{emcee} algorithm \citep{Foreman-Mackey13} and an affine-invariant ensemble sampler \citep{Goodman10}. The Bayesian framework uses the relative likelihood $\mathcal{L}$ at each step to determine the next move through the parameter space. The MCMC chain performs $3\times10^5$ calculations, for the \skysampler\ \citep{Smith19} or Gaussian approaches, respectively. The first 20\% of which are considered a burn-in phase and are excluded from the full analysis. The remaining 80\% of the iterations produce the final posterior distribution function (PDF) for the 13 free parameters. These standard options are implemented when running the \texttt{Python} wrapper code \texttt{KinMSpy\_MCMC}. For additional details on the implementation, we refer to \citet{Nguyen21, Nguyen22}.

For the \kinms\ modelling, we incorporate an additional mass term when calculating the gravitational potential. \citet{Boizelle21} estimate the total molecular gas mass of NGC~315 to be $M_\mathrm{gas} \approx 2.4\times 10^8$ \Msun, including corrections for helium. We assume that the gas mass follows the \cotwo\ surface brightness profile and include its gravitational contribution in \kinms\ via the \texttt{gasGrav} option, with the total gas mass treated as a free parameter. At the disk edge, the LOS speeds of $\sim$50 \kms\ due to the gaseous disk alone contribute minimally to the overall disk rotation.

\begin{figure*}[!th]
\centering
  \includegraphics[width=1.02\textwidth,trim=0cm 0cm 2cm 0cm,clip]{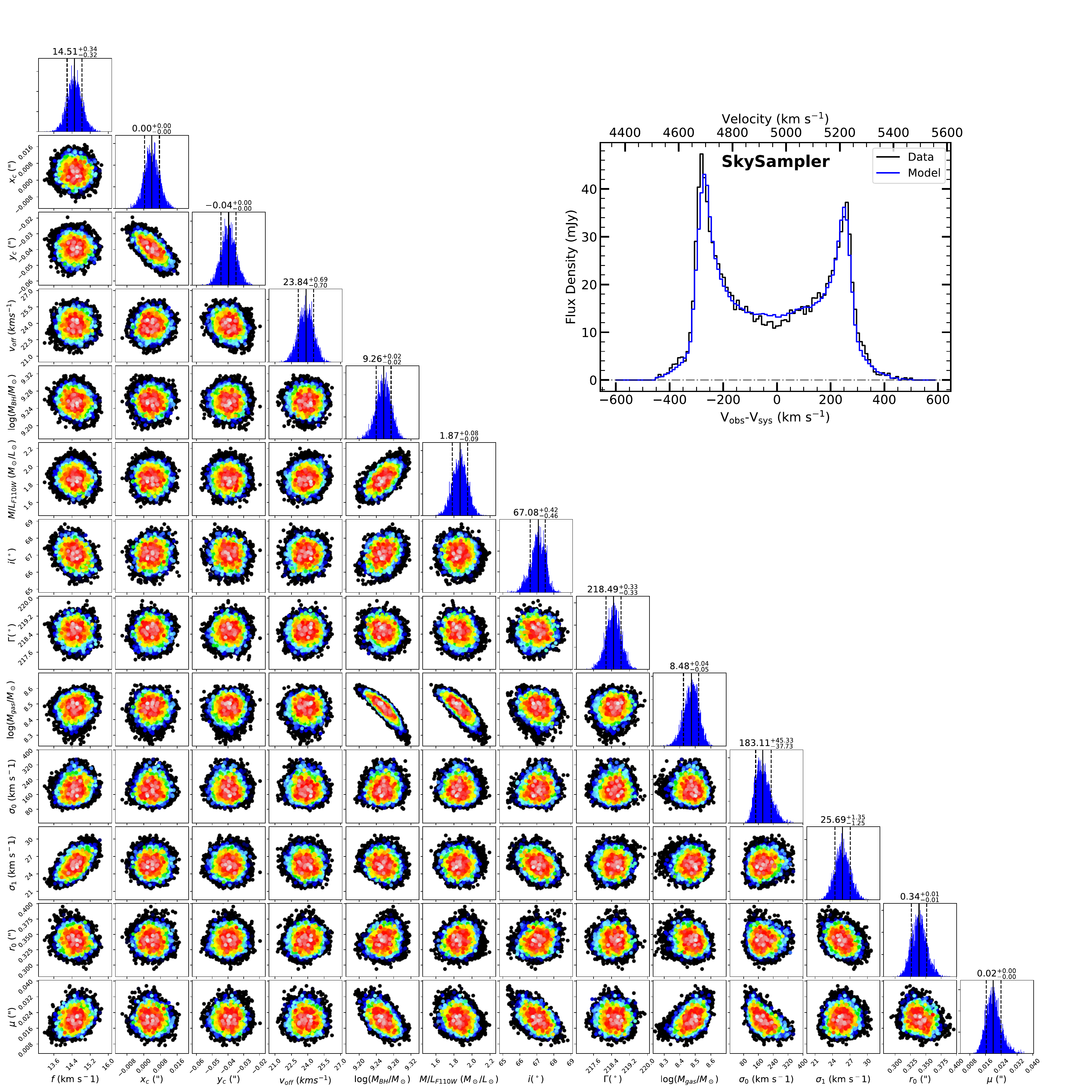}
  \caption{{\it Triangle:} Multi-dimensional projections of the posterior probability distributions from the \kinms\ modelling of the \cotwo\ datacube, in which, the \skysampler\ tool is used to describe the \cotwo\ surface brightness distribution. Additional details are provided in Section~\ref{kinms} and Figures~\ref{all_best_models} and \ref{all_bestmom1_models_reds}. The top panel of each column displays the marginalized PDF of the corresponding parameter, with the median indicated by the central solid line and the $1\sigma$ confidence interval by the outer dashed lines. Off-diagonal panels show the joint posterior distributions, where white, red, yellow-green, and black regions correspond to the $0.5\sigma$ (31--69 per cent), $1\sigma$ (16--84 per cent), $2\sigma$ (2.3--97.7 per cent), and $3\sigma$ (0.14--99.86 per cent) CLs, respectively. The $1\sigma$ and $3\sigma$ uncertainties for all fitted parameters are listed in Table~\ref{kimsbest}, and Figures~\ref{all_best_models} and \ref{all_bestmom1_models_reds} present a comparison between the observed and modeled \cotwo\ kinematics. Negative covariances between $M_{\rm BH}$ and $M/L_{\rm F110W}$ arise from the degeneracy between the gravitational potentials of the SMBH and the stellar component while there is a mild anti-correlation between $x_c$ and $y_c$ that reflects a model-fit degeneracy along the projected disk major axis. {\it Insert panel:} Comparison of the ALMA \cotwo\ integrated spectrum (panel F of Figure~\ref{comaps}) with the same profile extracted from the best-fitting model, which highlights the good agreement between the data and model.}
  \label{triangleSkysampler}   
\end{figure*}

\begin{figure*}[!th]
    \centering
    \includegraphics[width=1.02\textwidth,trim=0cm 0cm 2cm 0cm,clip]{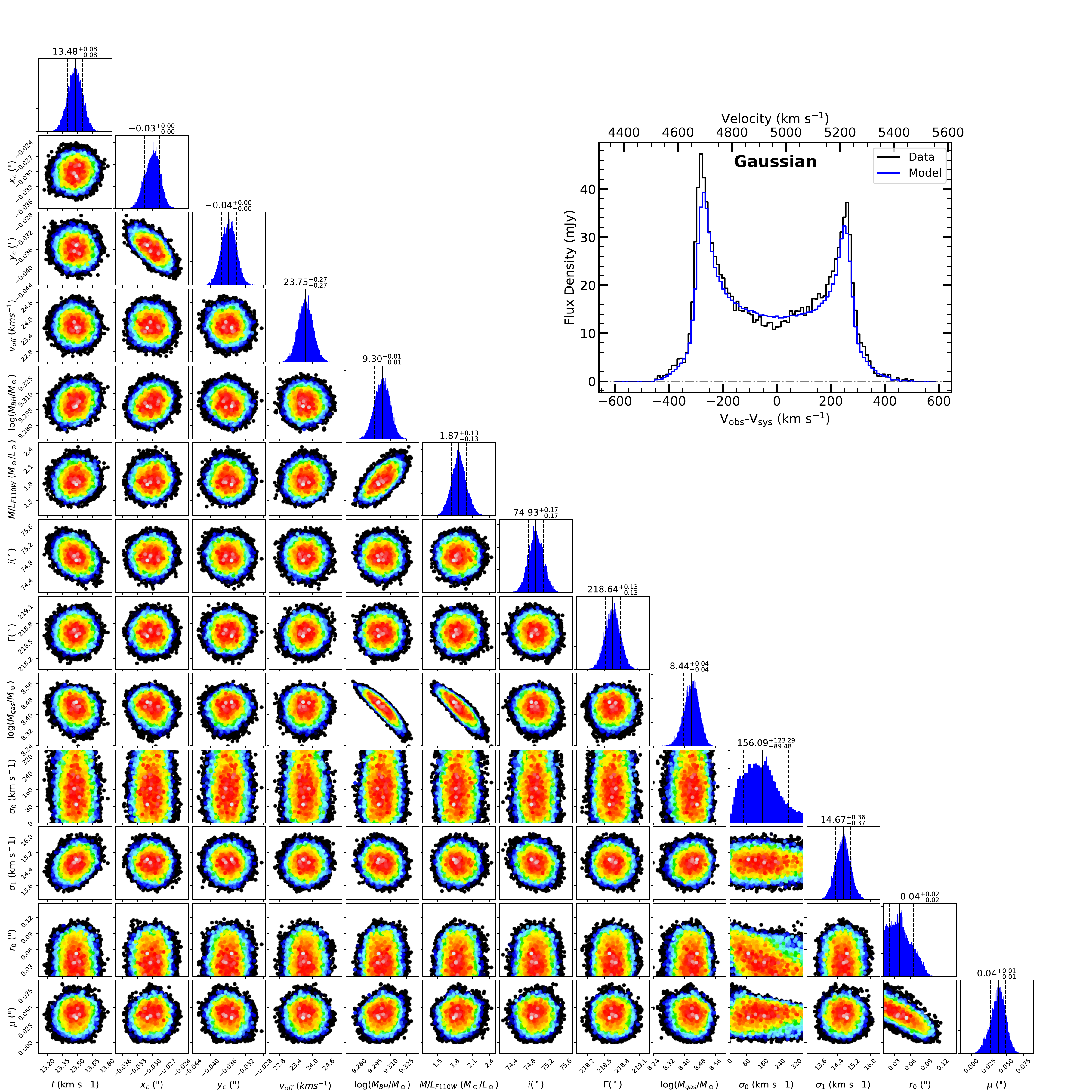}
  \caption{Similar to Figure~\ref{triangleSkysampler}, this figure presents the PDFs for the \kinms\ models of the \cotwo\ datacube, where a Gaussian profile describes the surface brightness distribution. Additional details are provided in Section~\ref{kinms} and Figures~\ref{all_best_models} and \ref{all_bestmom1_models_reds}. The $1\sigma$ and $3\sigma$ uncertainties for all fitted parameters are listed in Table~\ref{kimsbest}.}
  \label{triangleGauss}   
\end{figure*}

\begin{figure*}[!th]
  \centering
  \includegraphics[width=1.03\textwidth]{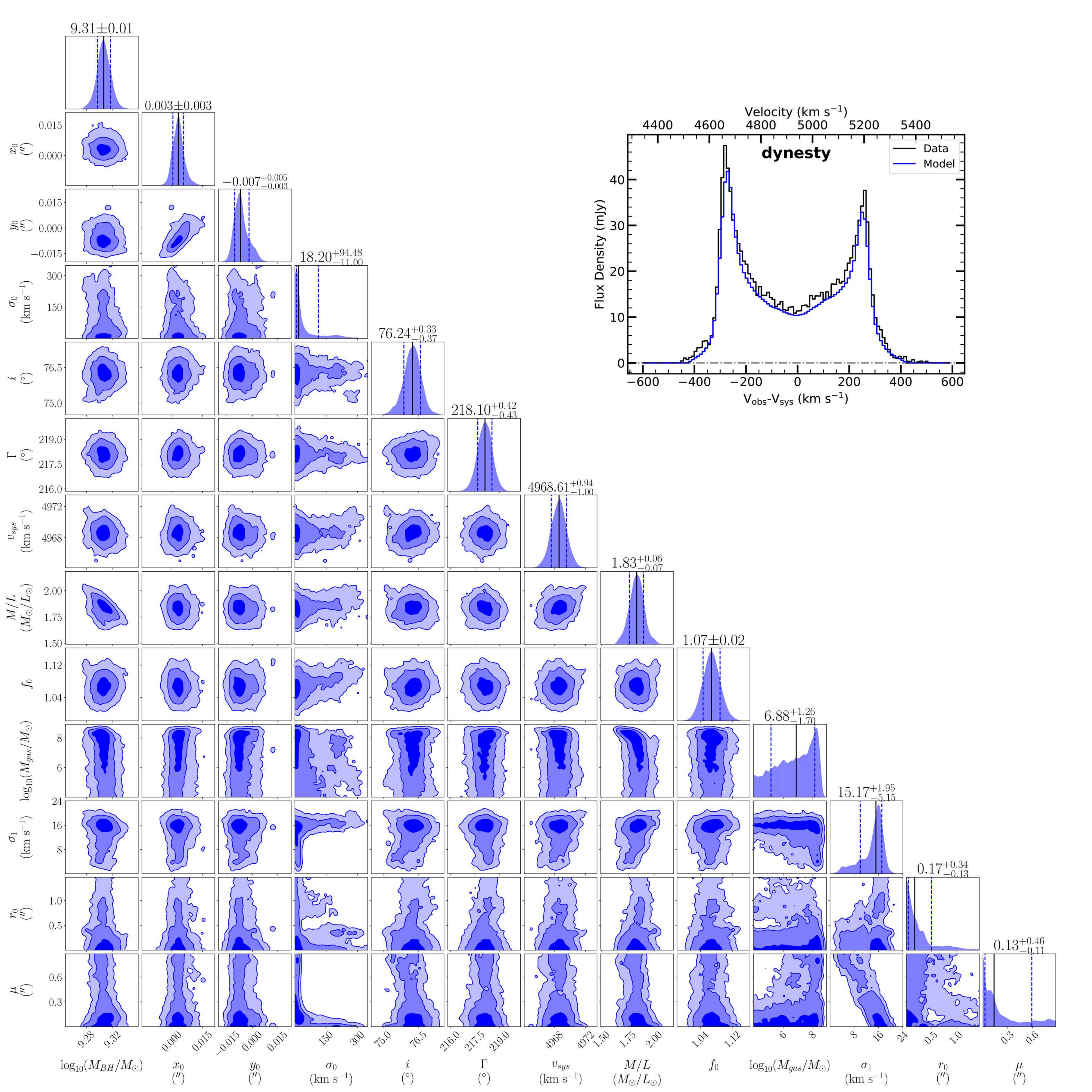}
  \caption{Similar to Figures~\ref{triangleSkysampler} and \ref{triangleGauss}, this figure shows the PDFs for the \dynesty\ modelling, which employs a deconvolved \cotwo\ flux map to weight the model line profiles. Additional details are provided in Section~\ref{dynestymodelling} and Figures~\ref{all_best_models} and \ref{all_bestmom1_models_reds}. The $1\sigma$ and $3\sigma$ uncertainties for all fitted parameters are listed in Table~\ref{kimsbest}. A similar, geometry-driven degeneracy of $x_c$ and $y_c$ is also present in the \dynesty\ modelling, with the opposite sign of the covariance due to an inverse x-axis coordinate projection.}
  \label{triangledynesty}   
\end{figure*}

\begin{table*}[!th]
    \caption{Gas-dynamical modelling results from the KinMS (gas distribution is described by a Gaussian or {\tt SkySampler}) and {\tt dynesty} modelling} 
    \centering
    \begin{tabular}{lcccc|ccc} 
    \hline\hline       
     Parameter & Search range & Best fit & $1\sigma$ Error& $3\sigma$ Error& Best fit & $1\sigma$ Error& $3\sigma$ Error\\
     &  &  & (16--84)\% & (0.14--99.86)\% &    & (16--84)\% & (0.14--99.86)\% \\  
     \multicolumn{1}{c}{(1)} & (2) & (3) & (4) & (5) & (3) & (4) & (5) \\  
    \hline 
     &  & \multicolumn{3}{c}{Gaussian} & \multicolumn{3}{c}{\tt SkySampler} \\ 
     \hline 
    {\it \underline{Mass Profile}:} &  &  &  &  &  &  \\
    $\log_{10}(M_{\rm BH}/{\rm M}_\odot)$&  $(8.5\to10)$ &  $9.30$ & $-0.01,+0.01$ & $-0.02,+0.02$  & $9.26$ & $-0.02,+0.02$ & $-0.06,+0.05$\\
    $M/L_{F\rm 110W}$               &  $(1\to2.5)$       &  $1.87$ & $-0.13,+0.13$ & $-0.38,+0.41$  & $1.87$ & $-0.09,+0.08$ & $-0.25,+0.25$\\
    $\log_{10}(M_{\rm gas}/{\rm M}_\odot)$&  $(8.5\to10)$&  $8.44$ & $-0.04,+0.04$ & $-0.13,+0.10$  & $8.48$ & $-0.05,+0.04$ & $-0.18,+0.13$\\[1mm]
    {\it \underline{Gas CND}:} & & & & & &   \\     
    $f$ (Jy~km~s$^{-1}$)            & $(9\to16)$    & $13.48$ & $-0.08,+0.08$ & $-0.23,+0.23$ & $14.51$ & $-0.32,+0.34$ & $-0.99,+1.02$\\
    $\Gamma$ ($\degr$)              &$(200\to240)$  & $218.64$ & $-0.13,+0.13$ & $-0.38,+0.39$ & $218.49$ & $-0.33,+0.33$ & $-1.01,+0.97$\\
    $i$ ($\degr$)                   & $(65\to85)$   & $74.93$ & $-0.17,+0.17$ & $-0.50,+0.49$ & $67.08$ & $-0.46,+0.42$ & $-1.67,+1.23$\\[1mm]
    {\it \underline{Nuisance}:} &  &  &  &  &  &  \\
    $\sigma_0$ (\kms)&$(0\to350)$    & $156.09$ & $-89.48,+123.29$ & unconstrained & $183.11$ & $-37.73,+45.33$ & $-89.24,+148.74$\\
    $\sigma_1$ (\kms)&$(0\to25)$    & $0.04$ & $-0.02,+0.02$ & $-0.03,+0.06$ & $25.69$ & $-1.25,+1.35$ & $-3.54,+4.07$\\
    $r_0$ ($\arcsec$)&$(0\to1.5)$   & $0.04$ & $-0.02,+0.02$ & unconstrained & $0.34$ & $-0.01,+0.01$ & $-0.03,+0.04$\\
    $\mu$ ($\arcsec$)&$(0\to0.9)$   & $0.04$ & $-0.01,+0.01$ & $-0.03,+0.03$ & $0.02$ & $-0.00,+0.00$ & $-0.01,+0.01$\\

    $x_{\rm c}$ ($\arcsec$)    &$(-0.2\to+0.2)$& $-0.030$ &$-0.001$, $+0.001$ & $-0.010$, $+0.010$ & $0.00$ & $-0.003$, $+0.004$ & $-0.010$, $+0.010$\\   
    $y_{\rm c}$ ($\arcsec$)    &$(-0.2\to+0.2)$& $-0.040$ &$-0.002$, $+0.001$ & $-0.010$, $+0.010$ & $-0.04$ & $-0.005$, $+0.004$ & $-0.010$, $+0.010$\\
    $v_{\rm off}$ (km~s$^{-1}$)&$(-70\to+130)$ & $23.70$ &$-0.19$, $+0.27$ & $-0.53$, $+0.72$ & $23.81$ & $-0.69$, $+0.68$ & $-1.90$, $+2.02$\\
    \hline
     &  & \multicolumn{3}{c}{\dynesty}\\  
     \hline 
    \textit{\underline{Mass Profile}:} &  &  &  &  \\
    $\log_{10}(M_{\rm BH}/{\rm M}_\odot)$ & $(8.5\to10)$  & $9.31$ & $-0.01,+0.01$ & $-0.03,+0.03$ &  &  &  \\
    $M/L_\mathrm{F110W}$ & $(1\to2.5)$  & $1.83$ & $-0.07,+0.06$ & $-0.19,+0.21$ &  &  &   \\
    $\log_{10}(M_{\rm gas}/{\rm M}_\odot)$ & $(4\to10)$  & $6.88$ & $-1.70,+1.26$ & $-2.87,+1.75$ &  &  &  \\[1mm]
    \textit{\underline{Gas CND}:} &  &  &  &  &  &  &   \\    
    $f$ (Jy~km~s$^{-1}$) & $(9\to16)$ & $12.50$ & $-0.25,+0.25$ & $-0.70,+0.75$ &  &  &   \\
    $\Gamma$ ($\degr$)  & $(200\to240)$ & $218.10$ & $-0.43,+0.42$ & $-1.43,+1.37$ &  &  &   \\
    $i$ ($\degr$) & $(65\to85)$ & $76.24$ & $-0.36,+0.33$ &  $-1.16,+1.06$ &  &  &   \\[1mm]
    \textit{\underline{Nuisance}:} &  &  &  &  &  &   \\
    
    $\sigma_0$ (\kms)   & ($5\to350$)  & 18.20     & $-11.00,+94.48$ & unconstrained &  &  &  \\
    $\sigma_1$ (\kms)   & ($5\to25$)   & 15.17      & $-5.15,+1.95$     & $-13.20,+5.60$ &  &  &  \\
    $r_0$ ($\arcsec$) & ($0\to1.50$) & 0.17  & $-0.13,+0.34$ & unconstrained &  &  &  \\
    $\mu$ ($\arcsec$) & ($0\to0.90$)    & 0.13     & $-0.11,+0.46$   & unconstrained &  &  &  \\
    
    $x_{\rm c}$ ($\arcsec$) & $(-0.06\to+0.06)$  & $0.003$ & $-0.003,+0.003$ & $-0.009,+0.012$ &  &  &   \\
    $y_{\rm c}$ ($\arcsec$) & $(-0.06\to+0.06)$ & $-0.007$ & $-0.003,+0.005$ & $-0.009,+0.019$ &  &  &   \\
    $v_{\rm off}$ (km~s$^{-1}$) & $(-70\to+130)$ & $26.51$ & $-1.00,+0.94$ & $-3.15$, $+3.05$ &  &  &  \\
     \hline
  \end{tabular}
 \parbox[t]{\textwidth}{{\bf Notes:} Best-fitting SMBH mass, stellar $M/L_{\rm F110W}$, and disk properties from Bayesian modelling of the ALMA \cotwo\ datacube of the NGC~315 CND. In these Bayesian approaches, \kinms\ is used first with model fluxes described by a Gaussian function and the \skysampler\ tool, followed by \dynesty\ nested sampling using a deconvolved CO moment map.  Columns list each parameter name (Column 1), search range (Column 2), best-fit values (Column 3; Figures \ref{all_best_models} and \ref{all_bestmom1_models_reds}), uncertainty at the $1\sigma$ CL ($16$--$84$ per cent of the PDF, Column 4) and $3\sigma$ CL ($0.14$--$99.86$ per cent of the PDF, Column 5). The $1\sigma$ CLs are also shown above each PDF in Figures~\ref{triangleGauss}, \ref{triangleSkysampler}, and \ref{triangledynesty}. The parameters $x_{\rm c}$, $y_{\rm c}$ and $v_{\rm off}$ are nuisance parameters defined relative to the adopted galaxy centre (R.A., Decl., $v_{\rm sys.}$)$\,=(0^{\rm h}57^{\rm m}48\fs85$, $+30\degr21\arcmin08\farcs80$, 4942.1 \kms).   For the \dynesty\ run, the velocity and flux axes are parameterized by the systemic velocity $v_{\rm sys}$ (barycentric) and a relative flux scaling $f_0$ of the deconvolved \cotwo\ moment-0 map, rather than by the velocity offset $v_{\rm off}$ and total flux $f$ used in the \kinms\ runs; the corresponding best-fit is $v_{\rm sys} = 4968.68^{+0.97}_{-1.02}$~\kms\ (equivalent to $v_{\rm off} \approx 26.6$~\kms) and $f_0 = 1.07\pm0.02$. The radial line-width parameters $r_0$ and $\mu$ of the \dynesty\ run are quoted in pc rather than arcsec. The central turbulent-dispersion parameters ($\sigma_0$, $r_0$, $\mu$) and the gas mass $M_{\rm gas}$ are only weakly constrained by the data and approach their prior bounds, reflecting the negligible dynamical leverage of the gas self-gravity and the compact central line-width component within the SMBH-dominated region; for these parameters we report only the $1\sigma$ interval.} 
  \label{kimsbest}
\end{table*} 

With the \kinms\ tool \citep{Davis13}, we followed two standard approaches to weight the model cube spectra. The first uses the \skysampler\ method \citep{Smith19}, which better reproduces the observed 2D \cotwo\ sky distribution; the second uses a parametric Gaussian function to describe the \cotwo\ flux distribution. We detail each approach and its best-fitting results in turn.

\vspace{0.2cm}
\noindent \textbf{(a) \skysampler\ Flux Map:} In an alternative fit, we model the \cotwo\ gas surface brightness using the {\skysampler}\footnote{\url{https://github.com/Mark-D-Smith/KinMS-skySampler}} tool \citep{Smith21} within the \kinms\ framework. The \skysampler\ tool constructs a molecular-cloud realization directly from the CLEAN components of the observed cube, thereby reproducing the intrinsic gas distribution and effectively isolating the kinematic modelling from assumptions about the analytic form of the surface brightness profile. In this implementation, the gas distribution is fixed by the CLEAN components and rescaled by a single free parameter, the total flux ($f$), since \skysampler\ assigns only relative cloud intensities. Because the CLEAN components exclude residual flux, their summed flux is slightly lower than that of the observed cube. The variable parameter $f$ mitigates some of this difference, although the restored missing flux is not always mapped to the location of the CO imaging residuals. All other aspects of the \kinms\ modelling, including the 13 free parameters listed in Table~\ref{kimsbest} and the fitting procedure, follow the general approach described in Section~\ref{kinms}.

\begin{figure*}
  \centering
  \includegraphics[width=0.9\textwidth]{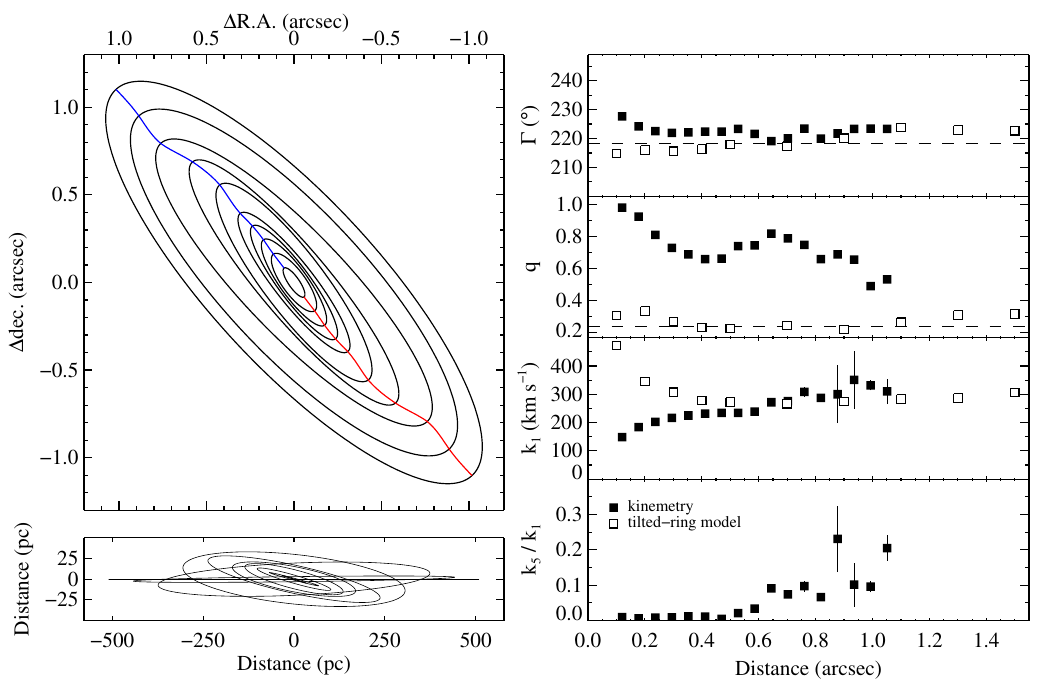}
  \caption{Disk structure modelling for the NGC~315 CND. From tilted-ring optimization (\text{left panels}), the best-fit CND structure shown both in projection (\textit{top}; with the line of nodes delineated) and in the plane of the outermost ring (\textit{bottom}; with $+x$ corresponding to the western direction) demonstrate that the warped disk remains relatively flat. \kinemetry\ results (\textit{right panels}) measured for most of the CND generally agree with the global kinematic position angle $\Gamma$ and $\cos i$ from \kinms\ modelling (dashed lines). The \kinemetry\ ellipse axis ratio $q$ remains elevated over the tilted-ring results due to beam-smearing effects that tend to circularize the axis ratios. The observed rotational speeds $k_1$ are lower than the best-fit $v_{\rm c} \sin(\cos^{-1} q)$ within about $2\times \overline{\theta}_{\rm FWHM}$ of the continuum center. This is similar to the results found by \citet{Boizelle19} that are attributed to beam smearing.}
  \label{ngc315_almac7m_kintwist}   
\end{figure*}

\vspace{0.2cm}
\noindent \textbf{(b) Gaussian Flux Profile:} This approach follows the same \kinms\ setup and free-parameter list as the \skysampler\ run described in Section~(a), differing only in that the \cotwo\ surface brightness is described by a parametric Gaussian rather than the CLEAN-component realization. At the physical resolution of our ALMA \cotwo\ data ($\sim$102 pc), the molecular gas surface brightness of NGC~315 is centrally peaked and fairly symmetrical. We found good overall agreement in preliminary \kinms\ fits using a single Gaussian, as shown in the one-dimensional (1D) profile extracted along the major axis in Figure~\ref{GaussianSB}. We note that this function does not fit the clumpy feature to the northeast or the general structure beyond $\sim$1\arcsec. Since these extended CO features are much beyond $R_{\rm SOI}$, they will not provide any useful constraint on the SMBH mass but should constrain the stellar $M/L_{\rm F110W}$ ratio. We optimized this analytic approximation to the \cotwo\ moment~0 map by convolving the model with the synthesized beam prior to fitting, finding a Gaussian dispersion $\mu_{\rm G}=0\farcs4$ ($\approx$136~pc) along the major axis. These parameters are fixed in the full \kinms\ modelling process fitting, while the total Gaussian flux ($f$) is left as a free parameter.

\subsubsection{\dynesty\ Modelling}\label{dynestymodelling}

To test the general consistency of the measured \Mbh\ results in a Bayesian framework, we also explored direct model cube fitting using the nested sampling code \dynesty\ \citep{Speagle20}. We adopted a standard likelihood $\mathcal{L} \propto \exp(-\chi^2/2)$ and used flat priors for the optimization process. We sampled \Mbh\ uniformly in logarithmic space while the remaining parameters are sampled uniformly in linear space. Unlike the \kinms\ approach, the \dynesty\ model uses the deconvolved \cotwo\ moment-0 map as the input CO surface brightness distribution, rather than the CLEAN-component-based \skysampler\ prescription. When running \dynesty, we used 250 live points to sample the prior, and the point with the lowest $\mathcal{L}$ is iteratively replaced with a new point at an improved likelihood. The threshold of 0.02 determines the stopping criteria and (initially) corresponds to the log-ratio of the current estimated Bayesian evidence and the evidence that remains to be sampled. For additional details about the implementation of this technique, we refer the reader to the discussion by \citet{Cohn21}.  

All 13 free parameters for standard \dynesty\ modeling runs are the same as those from \kinms\ efforts. For the molecular disc mass, we used the \cotwo\ surface brightness profile to estimate the corresponding circular velocities and scaled this profile by the total gas mass $M_\mathrm{gas}$.

\subsubsection{Results of Flat-disc Models}\label{results_flat_dics}

Figure~\ref{all_best_models} presents the PVDs for the best-fitting flat-disc models overlaid on our Cycle~7 \cotwo\ data for:
 
\begin{itemize}
		
	\item The best-fitting \kinms\ model in which the \cotwo\ surface brightness is parameterized by a Gaussian function. This model yields \Mbh~$= (2.00\pm0.05)\times10^9$ \Msun, $\ml_{\rm F110W} = 1.88^{+0.11}_{-0.10}$~\Msun/\Lsun, and $i=\left(74.99_{-0.79}^{+0.77}\right)\degr$, consistent with the values determined by \citet{Boizelle21} within their respective error budgets. The corresponding $\chi^2_{\rm red}\approx1.08$ indicates an overall good fit, although this is not unexpected given the large $N_{\rm KinMS}$, of which only a small fraction of pixels contain significant CO emission \citep[e.g.,][]{Boizelle19}.
	
	\item The best-fitting \kinms\ model in which the \cotwo\ gas distribution is generated using \skysampler. The best fit yields $M_{\rm BH}=(1.82\pm0.04)\times10^9$~M$_\odot$, $M/L_{\rm F110W}=1.87\pm0.08$~\Msun/\Lsun, and $i=\left(68.14_{-0.73}^{+0.68}\right)\degr$ has a final $\chi^2_{\rm red}\approx1.06$, indicating a marginal improvement relative to the model adopting the \textbf{Gaussian Flux Profile} for the \cotwo\ surface brightness. The derived \Mbh\ and $i$ values are both $\approx$9\% smaller than for the Gaussian model. $M/L_{\rm F110W}$ and the nuisance parameters remain nearly identical.
	
	\item The best-fitting \dynesty\ model, fitted over the same square spatial region corresponding to $N_{\rm KinMS}$ without block-averaging. The fit quality is likewise good, although the final $\chi^2_{\rm dof} \approx 1.10$ is marginally higher than for the \kinms\ results. These small differences likely arise from variations in the regions used to estimate the RMS noise. When sub-sampling and block-averaging the model and data cubes before fitting, we obtain $\chi^2_{\rm dof} \approx 1.66$. In all cases, reasonable choices of the $s$ and $d$ factors do not significantly affect the best-fit \Mbh\ value. Here, we reported the median posterior parameter values, including \Mbh~$= (2.02\pm0.05)\times 10^9$ \Msun, $\ml_{\rm F110W} = 1.83\pm 0.07$~\Msun/\Lsun, and $i=\left(76.24_{-0.36}^{+0.33}\right)\degr$. The \dynesty\ results are fully consistent with the \kinms\ Gaussian results within their $1\sigma$ CL. The modest offset relative to the \kinms\ \skysampler\ result amounts to $\sim$3$\sigma$ of the combined statistical uncertainties, reflecting the known sensitivity of \Mbh\ to the assumed CO surface brightness parameterization rather than a fundamental disagreement between the two modeling approaches.
    
\end{itemize}

Figure~\ref{all_bestmom1_models_reds} compares the observed mean LOS velocity map to best-fitting flat-disc models.  Overall, the best-fitting flat-disc models reproduce the observed mean LOS velocity field well. The residuals are generally small ($\lesssim$40~\kms) and localized, indicating that the full-cube modelling provides an adequate description of the full-disc CO kinematics. Some localized, spatially correlated residuals may suggest the presence of mild non-circular motions or radial flows. A more definitive assessment, however, would require higher physical-resolution CO imaging to characterize possible disk warping fully and to disentangle such effects from rotational broadening in this relatively edge-on disk.  
  
Here, all statistical uncertainties are quoted at the $1\sigma$ CL, corresponding to the 16th and 84th percentiles of the Bayesian PDFs. The corresponding posterior PDFs and corner plots are presented in Figure~\ref{triangleSkysampler} for the \kinms\ model using \skysampler, Figure~\ref{triangleGauss} for the best-fitting \kinms\ model with a Gaussian parameterization of the \cotwo\ surface brightness, and in Figure~\ref{triangledynesty} for the best-fitting \dynesty\ model. 

The SMBH mass and $M/L_{F\rm110W}$ measurements from flat-disc models are in good agreement with gas-dynamical modelling of Cycle~5 \cotwo\ imaging assuming the same underlying stellar mass model \citep{Boizelle21}. Bayesian formal errors on these two parameters are similar. However, the consistent statistical approach (and $N$ constraints) from PDFs results in smaller confidence intervals for other parameters in the \dynesty\ approach. 

These corner plots display the one-dimensional (1D) marginalized PDFs and 2D parameter covariances. All three flat-disc models, the two \kinms\ runs (\skysampler\ and Gaussian) and the \dynesty\ run, are optimized over the same set of 13 free parameters, differing only in the surface-brightness prescription, the treatment of $M_\mathrm{gas}$ velocity contributions, and in the parametrization of the velocity/frequency axis. Table~\ref{kimsbest} lists the best-fitting values of these parameters (primarily nuisance parameters beyond $M_{\rm BH}$ and $M/L$), together with the $3\sigma$ CL uncertainties (0.14th and 99.86th percentiles) and the adopted search ranges. Additional tests based on $\chi^2$ modelling and warped-disc models are discussed separately in Sections~\ref{additional} and \ref{tiltedmodel}, respectively.

Models that exclude either a central SMBH or the stellar mass contribution fail to reproduce the observed CO kinematics of the CND. In particular, the absence of an SMBH prevents the model from matching the inner, quasi-Keplerian rise of the velocity field, while neglecting the stellar potential leads to discrepancies across the more extended disk. Although increasing the stellar \ml$_{\rm F110W}$ can partially compensate for the absence of an SMBH and improve the fit to the inner kinematics, the values required ($\ml_{\rm F110W} > 10$~\Msun/\Lsun) are entirely inconsistent with predictions from old ($\sim$8--10~Gyr) single stellar population models \citep{vazdekis10}. We therefore conclude that no physically plausible model can simultaneously reproduce the full \cotwo\ kinematic structure without both the SMBH and the surrounding stellar mass.

As a consistency check on the dynamically inferred gas mass, we compare the best-fit $M_{\rm gas}$ from our Bayesian flat-disc models with the independent estimate derived from the integrated \cotwo\ flux via Equation~(1). The two \kinms\ runs recover $\log_{10}(M_{\rm gas}/M_\odot) = 8.46^{+0.03}_{-0.02}$ (Gaussian) and $8.47^{+0.04}_{-0.05}$ (\skysampler), corresponding to $M_{\rm gas} \approx (2.9$--$3.0)\times10^8\,M_\odot$. These dynamical values lie only $\approx$0.1~dex above the flux-based estimate of $M_{\rm gas} \approx 2.15\times10^8\,M_\odot$ (Section~\ref{momentmaps}) and the $2.4\times10^8\,M_\odot$ reported by \citet{Boizelle21}, well within the roughly factor-of-2 systematic uncertainty associated with the adopted CO-to-H$_2$ conversion factor \citep{Bolatto2013}. This agreement indicates that the gas self-gravity, incorporated through the \texttt{gasGrav} option, is captured at a level consistent with the observed CO luminosity. In the \dynesty\ run, by contrast, the best-fit $\log_{10}(M_{\rm gas}/M_\odot) = 6.88^{+1.26}_{-1.70}$ ($1\sigma$) is about 1.5 dex lower with a very broad (unconstrained) $3\sigma$ CL posterior. Given the consistent \Mbh\ and $M/L_{\rm F110W}$ values, we find the gas self-gravity within the SMBH-dominated region has negligible dynamical leverage. In the \kinms\ fits the recovered $M_{\rm gas}$ is negatively correlated with both $M_{\rm BH}$ and $M/L_{\rm F110W}$. For \dynesty\ runs, we find only a weak anti-correlation with these parameters. This confirms that treating $M_{\rm gas}$ and $\sigma_{\rm turb}$ as free parameters leaves \Mbh\ and its uncertainty essentially unchanged, while simultaneously validating the dynamical mass budget against the independent gas-mass determination.

\subsection{Frequentist $\chi^2$ Modelling Tests}\label{additional}

To test the impact of angular resolution on best-fitting parameter values, we fit model cubes directly to the data using straightforward $\chi^2$ optimization. These additional flat-disc models generally follow the iterative adjustment approach taken for Cycle 5 gas-dynamical modelling of NGC~315 \citep{Boizelle21}. None of these $\chi^2$ models include $M_\mathrm{gas}$ contributions. The baseline model~A assumes a radially uniform $\sigma_{\rm turb}$ and a dust-masked MGE ($A_J = 0$). Three models B1—B3 differ in the stellar MGE model, starting with dust-masked and then applying central $A_J=0.75,1.50$ mag dust correction of the stellar light behind the disk. Models B1--B3 all adopt a flexible Gaussian $\sigma_{\rm turb}$ profile.

In Table~\ref{tbl:chisq_results}, we report $\chi^2$ modelling results. There is little change between \Mbh\ of the other parameters between models A and B1 that only differ in the assumptions on $\sigma_{\rm turb}$. As expected, the very high but very compact Gaussian rise in intrinsic line widths has little impact on the goodness-of-fit value. Between models B1--B3, the results show a modest $\log(M_{\rm BH}$/\Msun) decline from $9.362$ to $9.280$ with increasing dust correction $A_J$, which are largely consistent with the same stellar luminosity model results for the Cycle~5 modelling \citep{Boizelle21}. The largest $M/L_{\rm F110W}$ change is the 6\% decrease from B1 to B2. These variations reflect the interplay between the recovered stellar light at the nucleus and the inferred central mass budget, further enhancing the existing degeneracy between \Mbh\ and $M/L_{\rm F110W}$. The best-fitting model B2 \Mbh~$= 2.046\times 10^9$ \Msun\ agree well with most Bayesian results.

\subsection{Warped-disc Models}\label{tiltedmodel}

In most of the CNDs probed using ALMA CO imaging, the molecular gas kinematics appear dynamically cold with $\sigma/v_c \ll 1$. Despite regular rotation, these moment 1 maps typically show mild to moderate kinematic twisting with $\Delta\Gamma \lesssim 30\degr$ across the disk extent \citep[in most cases; e.g.,][]{Boizelle17, Ruffa19}. Changes in inclination angle as a function of $R$ are less well constrained \citep[e.g.,][]{Boizelle19}. Even in cases with fairly small (observed) kinematic twists, the gas-dynamical modelling can be improved by adopting a more flexible model for the disk structure. \citet{Boizelle19} demonstrate the significant improvement in goodness-of-fit for the galaxy NGC~3258 (and a modest change in \Mbh) when adopting a tilted-ring model to describe an intrinsically warped disc solving for varying $\Gamma(R)$ and intrinsic axis ratio $q(R) \approx \cos i(R)$ for a thin disk.  Furthermore, they show that lower (physical) resolution CO imaging may hide kinematic warping, especially about the dynamic centre of the disc. Even if not detected at the current ALMA angular resolution, shifts in $\Gamma$ in the central few $\times$10~pc would be consistent with the molecular gas disk becoming progressively more aligned with the radio jet orientation at smaller scales, as recently observed in similar active galaxies \citep{Zhang2025b, Ruffa19}. For NGC~315, the CO moment~1 minor axis at a PA of $\sim -38\degr$ is already consistent with the radio jet orientation, whose estimated ${\rm PA}_{\rm jet}$ from VLBI observations varies between $\sim -47\degr$ and $-50\degr$ \citep{Boccardi21, Park21}. This suggests that the molecular gas and jet likely remain well aligned on smaller scales than those probed by our ALMA \cotwo\ data.

\vspace{0.2cm}
\noindent \textbf{(a) Non-uniform $\Gamma$ structure:} Following \citet{Nguyen20}, we characterize the kinematic structure of the NGC~315 CND using the \kinemetry\ code\footnote{\url{http://davor.krajnovic.org/idl/\#kinemetry}} \citep{Krajnovic06}, which generalizes surface photometry isophotal fitting techniques to the higher-order moments of the LOSVD. The method performs a harmonic expansion of the observed CO moment~1 map along best-fitting ellipses at successive radii, under the assumption that the velocity profile along each ellipse satisfies a simple cosine law consistent with circular rotation in a thin disk. At each radial annulus, the fit returns the kinematic position angle $\Gamma_k$, the ellipse axial ratio $q_k \approx \cos\,i$, and the harmonic coefficients of the expansion. The dominant term $k_1$ represents the amplitude of the bulk circular rotation, while the $k_5$ term captures deviations from pure circular motion, including contributions from non-circular flows, radial motions, and kinematic asymmetries. The ratio $k_5/k_1$ therefore serves as a diagnostic of the degree to which the observed velocity field departs from simple circular rotation, with $k_5/k_1 \lesssim 0.04$ generally considered consistent with axisymmetric circular motion \citep{Krajnovic06}.

Figure~\ref{ngc315_almac7m_kintwist} shows best-fit $\Gamma_k$, $q_k$, $k_1$, and $k_5$ values determined at 17 radial locations from the observed \cotwo\ moment~1 map. These solutions do not cover the entire CND extent due to higher noise and a rapid drop in covering fraction of moment~1 data points along ellipses near the disk edge. From $R \sim 0\farcs 1$, $\Gamma_k$ decreases from $\approx$228$^\circ$ and varies in a narrow $\approx$219--223$^\circ$ range for $0\farcs3 < R \lesssim 1\farcs1$, which is slightly elevated from the best-fit global $\Gamma$ values in Table~\ref{kimsbest}. \kinemetry\ $q_k$ solutions remain elevated over the best-fit $\cos i$ values from dynamical modelling due to beam smearing effects \citep{Boizelle19}. The ratio $k_5/k_1 \lesssim0.1$ across most of the radial locations does not suggest strong non-circular motions. We tested these $\Gamma_k$ solutions as input to the \kinms\ modelling, otherwise unchanged from the flat-disk approach in Section~\ref{kinms} using either a Gaussian function or \skysampler\ to characterize the \cotwo\ surface brightness. The best-fitting \Mbh~$= (1.94$--$1.98) \times 10^9$~\Msun\ is largely consistent with the uniform-$\Gamma$ results, with a modest improvement in $\chi^2_{\rm dof} \approx 1.05$, confirming that mild kinematic twists do not significantly bias the inferred \Mbh.

\begin{sidewaystable}
    \caption{Best-fitting parameters from frequentist implementation of the thin and warped-disk models.}
    \centering
    \begin{tabular}{cccccccccccccc}
        \hline\hline
        Model & Mass & $A_J$ & Disk & $\sigma_{\rm turb}$ & inflow &  &  &  &  &  &  &  &  \\
         & Model & (mag) & Structure &  &  &  &  &  &  &  &  &  &  \\
        \hline
        A  & MGE &  0   & flat & uniform & N &  &  &  &  &  &  &  &  \\
        B1 & MGE &  0   & flat & Gaussian & N &  &  &  &  &  &  &  &  \\
        B2 & MGE & 0.75 & flat & Gaussian & N &  &  &  &  &  &  &  &  \\
        B3 & MGE & 1.50 & flat & Gaussian & N &  &  &  &  &  &  &  &  \\
        C1 & MGE & 0.75 & warped & Gaussian & N &  &  &  &  &  &  &  &  \\
        C2 & $v_\mathrm{ext}$ &  & warped & Gaussian & N &  &  &  &  &  &  &  &  \\
        \hline 
        Model & $\log_{10}(M_{\rm BH})$ & $M/L_{\rm F110W}$ & $i$ & $\Gamma$ & $\sigma_1$ & $\sigma_0$ & $r_0$ & $\mu$ & $x_c$ & $y_c$ & $v_{\rm off}$ & $f_0$ & $\chi^2_{\rm dof}$ \\
         &(M$_\odot$)  & ($M_\odot/L_\odot$) & ($\degr$) & ($\degr$) & (\kms) & (\kms) & (pc) & (pc) & ($\arcsec$) & ($\arcsec$) & (\kms) &(Jy~\kms) & \\ 
        \hline
        A & 9.310 & 1.851 & 76.28 & 218.07 & 14.63 & -- & -- & -- & $-$0.0011 & $+$0.005 & 26.69 & 12.11 & 1.6612 \\
        B1 & 9.362 & 1.964 & 76.22 & 218.17 & 14.77 & 237.1 & 43.23 & 0.519 & $-$0.0005 & $+$0.0045 & 26.77 & 12.22 & 1.6759 \\
        B2 & 9.311 & 1.847 & 76.34 & 218.29 & 14.90 & 232.4 & 43.44 & 0.062 & $-$0.0011 & $+$0.0055 & 26.88 & 12.19 & 1.6619 \\
        B3 & 9.280 & 1.880 & 76.23 & 217.97 & 14.53 & 246.4 & 47.92 & $-$0.196 & $-$0.0028 & $+$0.0045 & 26.70 & 12.14 & 1.6702 \\
        C1 & 9.313 & 1.869 & -- & -- & 13.89 & 257.2 & 52.98 & $-$0.151 & $-$0.0013 & $+$0.0041  & 26.92 & 12.25 & 1.4772 \\
        C2 & 9.283 & -- & -- & -- & 13.89 & 257.2 & 52.98 & $-$0.151 & $-$0.0013 & $+$0.0041  & 26.81 & 12.25 & 1.4670 \\
    \hline
    \end{tabular}
    \parbox[t]{\textwidth}{\textit{Notes.} \textbf{Top:} Full-cube dynamical models and best-fitting parameters using $\chi^2$ minimization. MGE models are based on HST WFC3/F110W images after either dust masking ($A_J = 0$ mag) or dust correction ($A_J = 0.75$, 1.50 mag) of the stellar light behind the disk \citep{Boizelle21}. The major-axis position angle $\Gamma$ is measured east of north to the receding side of the disc. The spatial ($x_c$, $y_c$) and kinematic (\voff) centres are measured relative to the same continuum centroid and adopted recessional velocity as for Table~\ref{kimsbest}.}
    \label{tbl:chisq_results}
\end{sidewaystable}

\begin{figure*}[!th]
\centering
\includegraphics[width=0.8\textwidth]{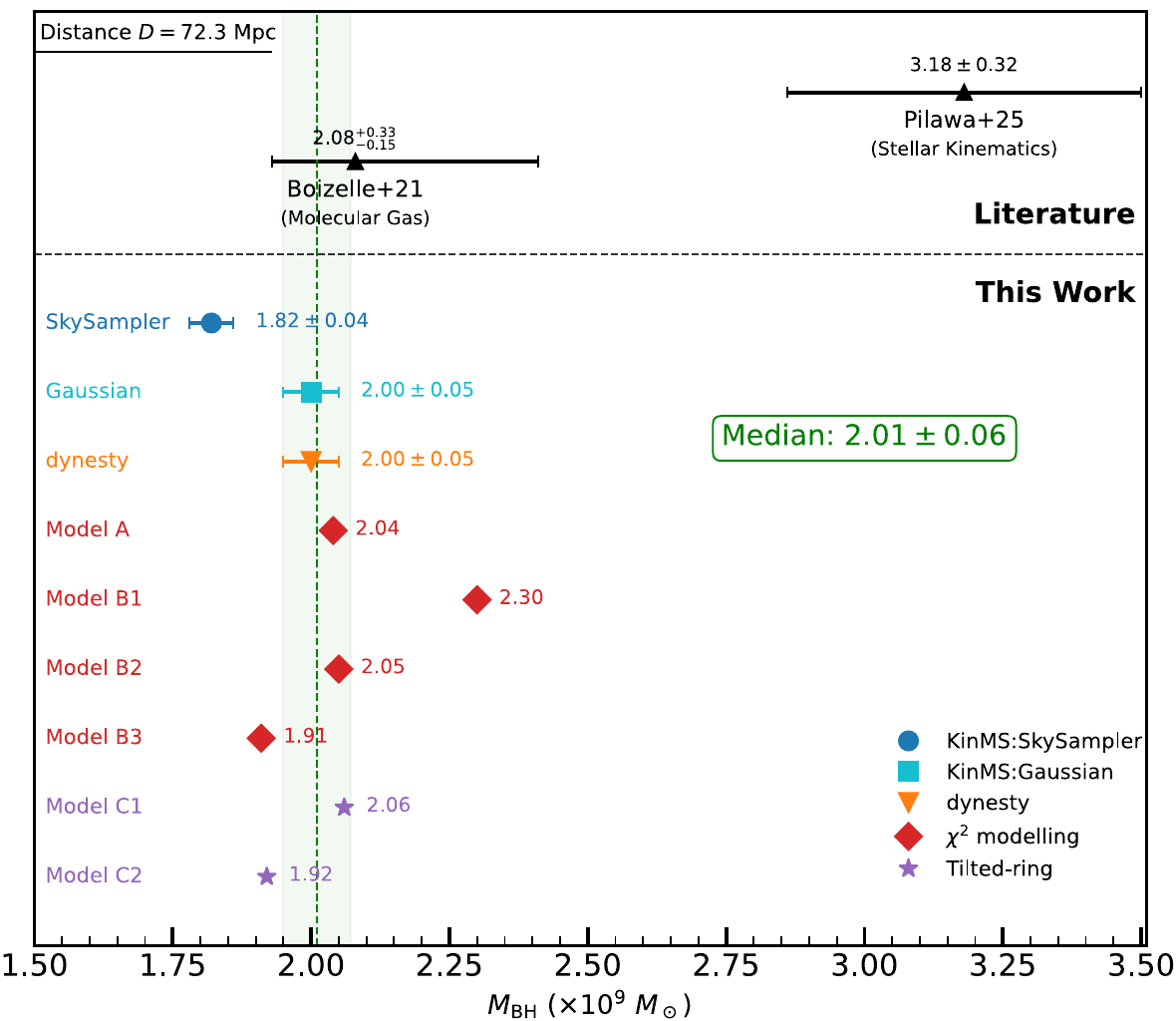}
\caption{\Mbh\ measurements for NGC~315 derived from our ALMA Cycle~7 observations using nine conventional molecular gas-based \Mbh\ models are summarized in Tables~\ref{kimsbest} and \ref{tbl:chisq_results}. The individual measurements and their $1\sigma$ uncertainties reflect variations arising from different assumptions about the molecular gas modelling. The vertical dashed lines indicate the adopted \Mbh, while the shaded regions denote the ensemble median and the 68\% bootstrap confidence interval as shown in the inserted panel. For comparison, we also show previous molecular gas-based estimates from \citet{Boizelle21}, obtained with lower-spatial-resolution ALMA Cycle~5 data, as well as the stellar-dynamical measurement from \citet{Pilawa2025} after adjusting to our adopted $D_L$.} 
\label{fig:NGC315_summary_all_models}
\end{figure*}

\vspace{0.2cm}
\noindent \textbf{(b) Tilted-ring Structure:} Next, we followed \citet{Boizelle19} in constructing and optimizing a model cube using the tilted-ring formalism to describe the disk structure. The optimization process no longer fits a global $\Gamma$ and $i$; instead, the disk structure is sampled at various ring locations  with individual $\Gamma$ and an assumed $q = \cos i$. At all locations, the tilted-ring model is still \textit{locally} thin and resides close to the galaxy's midplane. The model line-of-sight rotation field at each radius location is then computed as 
\begin{equation}
    v_\mathrm{LOS} = \vsys + v_\mathrm{c} \cos \Gamma \sin i \,.
\end{equation}
Here, we compute circular speeds $v_\mathrm{c}$ in the standard way using the input \Mbh\ and the same dust-corrected ($A_J = 0.75$ mag) MGE model and a free $M/L_{F\mathrm{110W}}$ as before. This tilted-ring model assumes a consistent \vsys\ at all radii but allows $\Gamma(R)$ and $q(R)$ values to vary with radius during the optimization process. 10 radial sampling locations were chosen to best sample the gradients seen in preliminary \kinemetry\ results while limiting the total optimization time. At intermediate radii between rings, the $\Gamma$ and $q$ values are determined by a linear interpolation.

\begin{figure*}[!th]
    \centering
    \includegraphics[width=0.9\textwidth]{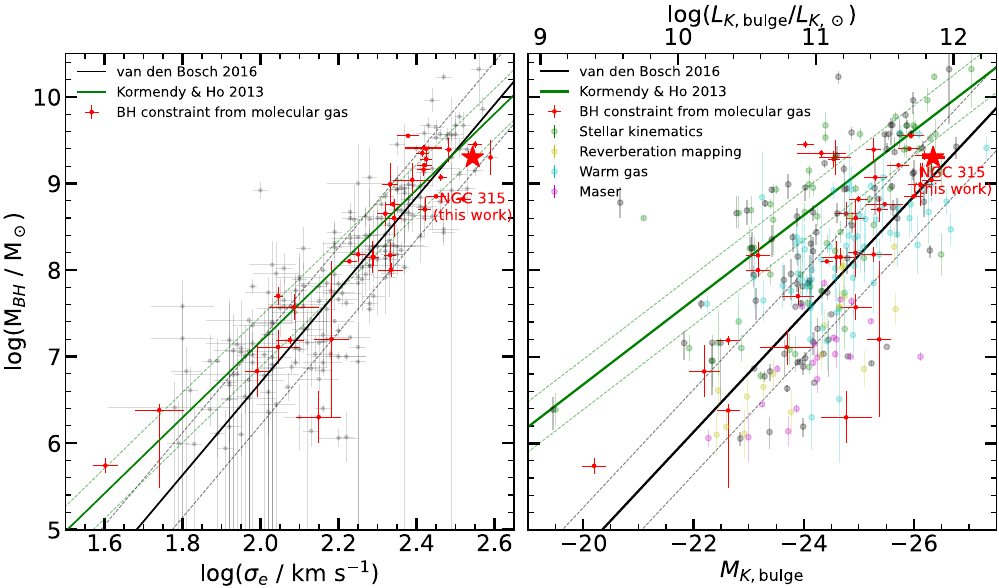} 
    \caption{The ensemble median value of our molecular-gas-based \Mbh\ measurements for NGC~315 is shown in comparison with the well-established literature \Mbh--$\sigma$ and \Mbh--$L_{K,\rm bulge}$ relations from \citet{Kormendy13} and \citet{vandenBosch2016}.} 
    \label{fig:bhmass-sigma}
\end{figure*}

The standard warped-disc approach (model C1) employs the same host galaxy luminosity as before, using the dust-corrected MGE and a free \ml\ ratio. A changing $i(R)$ does conflict with the assumption for the \texttt{mge\_circular\_velocity} task that tracers originate in the galaxy midplane, although the best-fitting $\Delta i(R) \approx 5\degr$ allows us to fix $i$ to previously-found values when calculating the stellar contributions to $v_\mathrm{c}$.

Introducing 20 (or more) free parameters to characterize the disk structure is time-consuming and increases the likelihood that the optimization will settle in local minima with unphysical $\Gamma(R)$ and $q(R)$ solutions. We therefore followed an iterative process that progressively improves the model cube goodness-of-fit while alternating between fixed disc structure with $R$ and fixed standard disc properties (e.g., \Mbh, \ml, etc.). Lastly, we ensured physical $\Gamma(R)$ and $q(R)$ solutions by forcing these values to smoothly vary from one ring location to the next with limits placed on deviations in $\Gamma$ and $q$ between adjacent rings. Best-fitting tilted-ring shape parameters are shown in Figure~\ref{ngc315_almac7m_kintwist}, and the final goodness-of-fit between models B2 and C1 improves from $\chi^2_{\rm dof} \sim 1.6619$ to $1.4772$.

Kinematic twists similar to those observed can also be introduced by radial flows, even in an otherwise flat disc in predominantly circular motion (I. Yoon 2017; B. D. Boizelle et al. 2019). As shown in Figure~\ref{ngc315_almac7m_kintwist}, the \kinemetry\ $k_5/k_1$ ratio---a standard diagnostic for non-circular motions---is almost identical to zero within $r<0\farcs5$ and reaches only the $\lesssim$0.1--0.2 level in the outer disc, indicating that radial flows contribute at most at the $\sim$10--20\% level relative to circular rotation. This outer-disc region carries little weight in constraining \Mbh. We therefore do not explicitly model bulk radial motion in our dynamical fits, as the small $k_5/k_1$ values suggest such motions are unlikely to significantly bias the inferred \Mbh.

Beyond these geometric considerations, residual mismatches between the model and
the observed kinematics remain even for the improved model~C1. The coarse, center-focused dust corrections to the F110W image and subsequent MGE fitting \citep{Boizelle21} is unlikely to recover the true nuclear stellar luminosity profile. In addition, we have not explored possible \ml\ gradients over the CND region \citep[e.g.,][]{Davis18}. Unlike the Bayesian \kinms\ and \dynesty\ fits of Sections~\ref{kinms} and \ref{dynestymodelling}, in which the gas self-gravity is included, the frequentist $\chi^2$ models A--C1 omit the gas-mass contribution to the circular velocity (Table~\ref{tbl:chisq_results}). Since the \cotwo\ data clearly show Keplerian-like rises in CO emission-line velocity towards the SMBH, we followed \citet{Boizelle19} in testing a simultaneous fit to both \Mbh\ and the extended stellar mass velocity contribution $v_\mathrm{ext}$, determined directly from the CO kinematics rather than fixed to the MGE. We incorporate this extended mass profile into model C2, while adopting the warped disk structure from model C1 for convenience.

\section{Discussion}\label{discussion}

\subsection{Ensemble Median-\Mbh\ Value and Error Budgets}\label{error_budgets}

Across the nine molecular gas dynamical models explored in this work, the spread in inferred \Mbh\ values of $(1.91–2.51)\times10^9$~\Msun\ primarily reflects differences in model setup rather than statistical uncertainties. We thus determine the final \Mbh\ value and error budget for NGC~315 using an ensemble modelling approach based on nine molecular gas dynamical models, extending beyond the uncertainties captured by any single Bayesian posterior. Specifically, we construct 9 independent dynamical models (Tables~\ref{kimsbest} and \ref{tbl:chisq_results}) that adopt different physical assumptions within the \kinms, \dynesty, and $\chi^2$ frameworks, including tilted-ring and $v_\mathrm{ext}$ additions. The \Mbh\ measurements and their $1\sigma$ uncertainties are summarized in Figure~\ref{fig:NGC315_summary_all_models}, providing a consolidated view of the constraints derived from the spatially resolved ALMA Cycle~7 \cotwo\ kinematics. This ensemble strategy quantifies the impact of methodological and physical assumptions on the inferred mass \citep{Nguyen_2026_M81, Nguyen_2026_4061}. We estimate the final \Mbh\ by bootstrap resampling these nine measurements to incorporate inter-model variance into the uncertainty budget. Adopting the median and the 68\% bootstrap CL ($1\sigma$), we obtain \Mbh$/10^9$~\Msun$ = 2.02^{+0.04}_{-0.05}{\rm(stat)}^{+0.05}_{-0.04}{\rm(sys)}$. Given the 3.7\% uncertainty in $D_L$ \citep[from surface brightness fluctuation measurements of the F110W data;][]{Jensen21}, the absolute \Mbh$/10^9$~\Msun\ precision for NGC~315 would include an additional $\pm 0.07$ (dist) term.

Our new molecular-gas-based \Mbh\ measurement from the ALMA Cycle~7 data for NGC~315 is 3\% higher than that derived from the lower-spatial-resolution Cycle~5 observations \citep{Boizelle21}, and 26\% and 32\% lower than the value predicted by the \citet{Kormendy13} \Mbh--$\sigma$ relation for bulge galaxies and the stellar-dynamical measurement from Gemini and McDonald spectroscopy \citep{Pilawa2025}, respectively.

As discussed in \citet{Boizelle21}, the dominant systematic uncertainty in their NGC~315 measurement arises from uncertainties in the luminous mass model due to the presence of nuclear dust. Across the three dust-masked or dust-corrected MGEs their \Mbh\ measurements vary by $\approx$21\%. Our Cycle~7 higher angular resolution CO(2-1) imaging better isolates the locus of rapid gas rotation within $R_{\rm SOI}$, further limiting residual degeneracy between the SMBH and the central stellar mass distribution, resulting in a smaller $\approx$18\% spread between models B1--B3. Because the stellar mass profile is regularized by the MGE with a spatially constant $M/L$, this improved constraint on the central mass distribution propagates outward, indirectly reducing the uncertainty on the extended stellar mass distribution as well. We expect that CO imaging with a similar line sensitivity of $\approx$70~$\mu$Jy~beam$^{-1}$ per 10~\kms\ channel and an angular resolution of $0''.1$ would further tighten these constraints, enabling percent-level precision on \Mbh\ with a more complete exploration of the extended stellar mass distribution to better understand discrepancies with results from \citet{Pilawa2025}.

Even for the \kinms\ models, the best-fit $\chi^2_\mathrm{dof}$ values are \textit{formally} unacceptable while still providing close agreement with the observed CND kinematics (see Figure~\ref{all_best_models} and \ref{all_bestmom1_models_reds}). However, part of this apparent discrepancy arises from an imperfect treatment of noise that is correlated on beam scales \citep[see the discussion by][]{Davis17}. Downsampling in the spatial dimension mitigates the impact of correlated noise for $\chi^2$ and (some of) the \dynesty\ modelling runs, although the overall effect on the best-fit parameter values is small.

\subsection{Resolving the Black Hole Sphere of Influence}\label{soi} 

Our improved mass estimate found that $R_{\rm SOI}\approx0\farcs8$ with \cotwo\ emission detected down to $\approx$$0.11R_{\rm SOI}$, corresponding to the criterion for accurately measuring the mass of a SMBH based on the resolved power of radio observations $\zeta=2R_{\rm SOI}/\theta_{\rm beam}\approx7.6$ \citep{Boizelle19, Boizelle21, Nguyen20}, implying that our higher angular resolution ALMA observation is totally resolved with $\zeta\cos i\approx2.5$. Our dynamically derived \Mbh\ for NGC~315 thus meets the above criterion of precision.

\subsection{Comparison Between Gas- and Stellar-Dynamical \Mbh\ Measurements}\label{bh_correlations}

As shown in Figure~\ref{fig:bhmass-sigma}, our improved molecular-gas-based \Mbh\ for NGC~315 from the Cycle~7 ALMA \cotwo\ data at $\overline{\theta}_{\rm FWHM}\approx 0\farcs2$ is consistent with the value predicted from the stellar velocity dispersion and $K$-band luminosity of massive elliptical and bulge galaxies \citep{Kormendy13}. It is also consistent, within the intrinsic scatter, with the \citet{McConnell13} and \citet{Saglia16} scaling relations. Details of the stellar bulge mass and bulge $K$-band luminosity of NGC~315, along with their corresponding predicted \Mbh\ values, can be found in the discussion section of \citet{Boizelle21}.
 
Our ensemble median-\Mbh\ estimate for the SMBH in NGC~315, derived from three independent dynamical modelling approaches (using both Bayesian and frequentist statistics) is consistent within $1\sigma$ statistical uncertainty with previous measurements based on the same transition from the lower-resolution ALMA Cycle~5 data ($\overline{\theta}_\mathrm{FWHM} \approx 0.3\arcsec$). However, our value remains 32\% smaller than the recent stellar-dynamical measurement obtained from high signal-to-noise, $0.3\arcsec$ resolution integral-field spectroscopy at the Gemini and McDonald Observatories \citep{Pilawa2025}. Future higher-spatial-resolution JWST observations may help reduce this discrepancy, or at least allow for a more direct comparison between precision ALMA CO-based \Mbh\ measurements and stellar-dynamical determinations to better understand (and perhaps refine) the underlying assumptions in both approaches.

For a larger set of dynamical \Mbh\ cross-checks, there exists a factor of $\sim$2 systematic offset toward higher \Mbh\ values from stellar-dynamical measurements compared to gas-dynamical determinations \citep[see figure~2 of][]{Thater19}. However, it is important to distinguish between the two gas-dynamical tracers, as they behave differently in cross-comparisons. Ionized-gas measurements frequently disagree with stellar-dynamical results, with discrepancies often attributed to substantial non-circular and planar motions driven by radial flows, turbulent pressure support \citep{Barth01}, or interactions with radio jets \citep{VerdoesKleijn06} that are especially common in massive elliptical galaxies \citep{Noel-Storr03, Noel-Storr07}. Incorporating non-Keplerian motions into ionized-gas dynamical modelling has been shown to increase the best-fit \Mbh\ by a factor of $\sim$2 \citep{Jeter19}.

By contrast, cold molecular gas dynamical measurements are generally in good agreement with stellar-dynamical results, as physically and dynamically cold CO-bright discs show little support for turbulent pressure or significant radial flows \citep[e.g.,][]{Boizelle19, Cohn21}. Notable exceptions include NGC~1332 \citep{Rusli11, Barth16} and NGC~315 \citep{Boizelle21}, the subject of this work, where molecular gas and stellar dynamical measurements disagree at the factor of $\sim$1.5 level. In the case of NGC~1332, the discrepancy may reflect systematic uncertainties intrinsic to the stellar dynamical method itself, including assumptions about the stellar $M/L$, orbital anisotropy, and the construction of the MGE stellar luminosity model \citep{Davis20, Dominiak2025}. For NGC~315, a settled gaseous disc with circular orbits may not be a consistent assumption given its mildly triaxial potential \citep{Pilawa2025}. The higher-resolution ALMA \cotwo\ observations of NGC 315 presented here confirms the tension with stellar-dynamical results is not due to the specific gas-dynamical modelling approach.

\section{Conclusions}\label{conclusion}

We present ALMA Cycle~7 \cotwo\ observations of NGC~315 at an angular resolution of $0\farcs230\times0\farcs175$ and use these data to obtain a high-precision measurement of its central SMBH mass while systematically comparing multiple gas-dynamical modelling approaches. Our principal conclusions are summarized as follows:

\begin{enumerate}

\item The ALMA data resolve the SMBH SOI by a factor of $\approx$7.6 across the beam FWHM, providing strong dynamical leverage on the Keplerian rise of the \cotwo\ CND and substantially reducing degeneracies between \Mbh\ and the stellar $M/L_{\rm F110W}$ ratio.

\item A suite of full-cube forward models yield molecular-gas-based SMBH masses in the range $(1.8$--$2.3)\times10^9$~\Msun. We directly compare results from \kinms, \dynesty\ (nested sampling), and frequentist ($\chi^2$) approaches when applied to the same dataset. Most runs produce consistent \Mbh\ values within their statistical uncertainties, demonstrating that when the SOI is well resolved, the inferred mass is not strongly method-dependent. The largest contributions to the \Mbh\ error budget stem from assumptions regarding the stellar mass model (e.g., MGE parameterization and dust treatment), the gas surface-brightness distribution (Gaussian versus \skysampler), and disk geometry (flat versus warped).

\item Tests incorporating tilted-ring geometries do not significantly change the best-fit $M_{\rm BH}$, indicating that mild kinematic twists or departures from a strictly flat disk do not drive the mass determination. Although bulk radial motions are not explicitly modelled, the typical \kinemetry\ $k_5/k_1\lesssim 0.1$ values across most of the CND suggest their contributions remain modest.

\item Combining the ensemble of independent molecular-gas dynamical models, we derive an ensemble median SMBH mass of $M_{\rm BH}/\times10^9 \,M_\odot = 2.02^{+0.04}_{-0.05}(\mathrm{stat})^{+0.05}_{-0.04}(\mathrm{sys})$ that is in agreement with established \Mbh$-\sigma_\star$ and \Mbh$-L_{\rm bulge}$ scaling relations. The first quoted uncertainty reflects statistical and random components, while the second encompasses systematic effects added in quadrature that define the true precision floor.

\item NGC~315 is a rare system with cross-validated SMBH measurements from multiple techniques, making it a benchmark system for quantifying methodological systematics in molecular gas-based SMBH measurements. Our molecular-gas-based \Mbh\ is consistent with previous ALMA measurements but 32\% lower than independent, ground-based stellar-dynamical determinations at similar resolution. Higher-resolution ALMA \cotwo\ observations and JWST stellar kinematic measurements would further constrain \Mbh\ and reduce the systematic uncertainties identified in this work.

\end{enumerate}

Given its large SMBH mass, well-resolved SOI, and radio-loud nucleus, NGC~315 is also a promising target for accretion studies. Future ultra-high-angular-resolution facilities such as the Black Hole Explorer \citep[BHEX;][]{Hudson2025}, capable of achieving sub-10~$\mu$as resolution, may enable horizon-scale imaging of the jet-launching region and the SMBH shadow in NGC~315, beyond the reach of current ground-based Event Horizon Telescope arrays.

\section*{Acknowledgements}
The authors would like to thank the anonymous referee for their careful reading and useful comments, which helped to improve the paper greatly.  D.D.N acknowledges support from the ELT postdoctoral research fellowship at the Department of Astronomy, University of Michigan. B.D.B acknowledges access to the BYU Office of Research Computing HPC resources. Research at Brigham Young University is also support by the College of Computational, Mathematical, and Physical Sciences.  S.T. acknowledges funding from the European Research Council (ERC) under the European Union's Horizon 2020 research and innovation programme under grant agreement No 724857 (Consolidator Grant ArcheoDyn).

This paper makes use of the following ALMA data: ADS/JAO.ALMA\#2019.1.00036.S. ALMA is a partnership of ESO (representing its member states), NSF (USA), and NINS (Japan), together with NRC (Canada), NSC and ASIAA (Taiwan), and KASI (Republic of Korea), in cooperation with the Republic of Chile. The Joint ALMA Observatory is operated by ESO, AUI/NRAO, and NAOJ. The National Radio Astronomy Observatory is a facility of the National Science Foundation operated under cooperative agreement by Associated Universities, Inc. The work has made use of the NASA/IPAC InfraRed Science Archive, which is funded by NASA and operated by the California Institute of Technology.

\facility{ALMA.}

\software{{\tt Python~v3.12} \citep{VanRossum2009}, 
{\tt Matplotlib~v3.6} \citep{Hunter2007}, 
{\tt NumPy~v1.22} \citep{Harris2020}, 
{\tt SciPy~v1.3} \citep{Virtanen2020},  
{\tt photutils~v0.7} \citep{bradley2024}, 
{\tt AstroPy~v5.1} \citep{AstropyCollaboration2022}, 
{\tt AdaMet v2.0} \citep{Cappellari2013a}, 
{\tt JamPy~v7.2} \citep{Cappellari2020}, 
{\tt pPXF~v8.2} \citep{Cappellari23}, 
{\tt vorbin~v3.1} \citep{Cappellari2003}, and
{\tt MgeFit~v5.0} \citep{Cappellari02}.
} 

\bibliographystyle{aasjournalv7}
\bibliography{ngc315}

\label{lastpage}
\end{document}